\documentclass[twocolumn]{aa} 

\usepackage[varg]{txfonts}
\usepackage{natbib}
\bibpunct{(}{)}{;}{a}{}{,} 
\usepackage{graphicx,tabularx,stmaryrd,amsmath,multirow}
\usepackage[]{hyperref}
\begin{document}

\title{Chemistry in disks}

\subtitle{XI. Sulfur-bearing species as tracers of protoplanetary disk physics and chemistry: the
DM~Tau case.}

 \author{D. Semenov\inst{1,2}
  \and C. Favre\inst{3}
  \and D. Fedele\inst{3}
  \and S. Guilloteau\inst{4,5}
  \and R. Teague\inst{6}
  \and Th. Henning\inst{1}
  \and A. Dutrey\inst{4,5}
  \and E. Chapillon\inst{7}
  \and F. Hersant\inst{4,5}
  \and V. Pi\'etu\inst{7}
}
\institute{Max-Planck-Institut f\"{u}r Astronomie, K\"{o}nigstuhl 17, D-69117 Heidelberg, Germany\\
\email{semenov@mpia.de}
\and Department of Chemistry, Ludwig Maximilian University, Butenandtstr. 5-13, D-81377 Munich, Germany
\and INAF, Osservatorio Astrofisico di Arcetri, L.go E. Fermi 5, I-50125 Firenze, Italy
\and LAB, Universit\'e de Bordeaux, B18N, All\'ee Geoffroy, Saint-Hilaire, CS 50023, 33615 Pessac Cedex
\and CNRS, Universit\'e de Bordeaux, B18N, All\'ee Geoffroy, Saint-Hilaire, CS 50023, 33615 Pessac Cedex
\and Department of Astronomy, University of Michigan, 1085 S. University Avenue, Ann Arbor, MI 48109, USA
\and IRAM, 300 Rue de la Piscine, F-38046 Saint Martin d'H\`{e}res, France
}

\titlerunning{Chemistry in Disks. XI}
\authorrunning{D.\ Semenov\ et al.}
   \date{}


  \abstract
   {Several sulfur-bearing molecules are observed in the interstellar medium and in comets, in strong contrast to protoplanetary disks where only CS,
   H$_{2}$CS and SO have been detected so far. }
   {We combine observations and chemical models to constrain the sulfur abundances and their sensitivity to physical and chemical conditions in the DM~Tau protoplanetary disk.}
   {We obtained $0.5\arcsec$ ALMA observations of DM~Tau in Bands 4 and 6 in lines of CS, SO, SO$_{2}$, OCS, CCS, H$_{2}$CS and H$_{2}$S, achieving a $\sim 5$~mJy sensitivity. Using the non-LTE radiative transfer code RADEX and the forward-modeling tool DiskFit, disk-averaged CS column densities and upper limits for the other species were derived.}
   {Only CS was detected with a derived column density of $\sim 2-6\times 10^{12}$~cm$^{-2}$. We report a first tentative detection of SO$_{2}$ in DM~Tau.
   The upper limits range between $\sim 10^{11}$ and $10^{14}$~cm$^{-2}$ for the other S-bearing species.
   The best-fit chemical model matching these values requires a gas-phase C/O ratio of $\gtrsim 1$ at $r\gtrsim 50-100$~au.
   With chemical modeling we demonstrate that sulfur-bearing species could be robust tracers of the gas-phase C/O ratio,
   surface reaction rates, grain size and UV intensities.
   }
   {The lack of detections of a variety of sulfur-bearing molecules in DM~Tau other than CS implies a dearth of reactive sulfur in the gas phase, either through efficient freeze-out or because most of the elemental sulfur is in other large species, as found in comets. The inferred high CS/SO and CS/SO$_{2}$ ratios require a non-solar C/O gas-phase ratio of $\gtrsim 1$, consistent with the recent observations of hydrocarbon rings in DM~Tau. The stronger depletion of oxygen-bearing S-species compared to CS is likely linked to the low observed abundances of gaseous water in DM~Tau and points to a removal mechanism of oxygen from the gas.
 }


   \maketitle
%

\section{Introduction}
\label{sec:intro}
One of the most exciting topics in astronomy is to understand the
physical and chemical link between various stages of the star and planet formation processes
and the properties of planetary systems and exoplanet atmospheres \citep[e.g.,][]{2012A&A...541A..97M,Molliere_ea15a,Cridland_ea17}.
Infrared and (sub-)millimeter observations of the dust continuum and molecular lines are used to study
these environments. The compact, $\la 100-1\,000$~au sizes and the low, $\la 0.01-0.1M_{\sun}$ gas
masses make such studies particularly challenging for protoplanetary disks,
even with the most powerful observational facilities like ALMA, NOEMA, and the EVLA.
Thus, compared to more than 200 molecules discovered in the interstellar medium\footnote{http://www.astro.uni-koeln.de/cdms/molecules},
only $\sim 20$ species and their isotopologues were detected in disks, see e.g. \citet{2013ChRv..113.9016H,2014prpl.conf..317D,Bergin:2015in}.


Using this limited set of molecular tracers, crucial information about disk physics, chemistry and kinematics at outer radii of
$\gtrsim 50-100$~au have been obtained.
From the relatively easy-to-detect CO isotopologues disk orientations, sizes, temperature structures, and indirect disk gas masses have been derived
\citep[e.g.,][]{Williams_Best14,Ansdell:2017tr,Ansdell:2016gm,2016A&A...592A.124G,Pascucci:2016ff,Schwarz:2016ch,Long:2017wc}.
More reliable gas mass measurements using HD have been obtained with {\em Herschel}
for TW~Hya, DM~Tau, and GM~Aur \citep{Bergin_ea13,McClure_ea16,Trapman:2017tn}.
The importance of CO-regulated ice chemistry for the synthesis of complex organic molecules has been supported by the
H$_{2}$CO observations of DM~Tau \citep{Loomis:2015es} and CH$_{3}$OH observations of TW~Hya \citep{Walsh2016,Parfenov2017}.

One of the most intriguing ALMA results is evidence of substantial depletion of
elemental C and O in T~Tauri disks, with C/O becoming $>1$ in a molecular layer \citep{Bergin:2016ge,Oberg:2016fh}.
\citet{Schwarz:2016ch} and \citet{Kama_ea16} have found that CO is depleted by two orders of magnitude
in the TW~Hya disk, which cannot be solely explained by the CO freeze-out in the midplane,
while CO depletion in a warmer Herbig~Ae disk HD~100546 is negligible \citep{Kama_ea16}.
Similarly, \citet{McClure_ea16} derived CO depletion of up to a factor of 5 for DM~Tau
and up to a factor of $\sim 100$ for GM~Aur.
Plausible mechanisms for this depletion are: 1) warm surface chemistry that converts volatile CO to less volatile CO$_{2}$,
2) He$^{+}$ destruction of CO, which liberates carbon that is converted into refractory complex organics, or
3) preferential removal of oxygen from the disk molecular layer by
the freeze-out of water onto sedimenting large dust grains \citep[e.g.][]{SW2011,2013ApJ...776L..38F,Krijt_ea16,Oberg:2016fh,Rab_ea17}.
The higher freeze-out temperature of
water, $T_{\rm fr}\sim 140-160$~K, favors its depletion over the volatile CO ($T_{\rm fr} \sim 20-25$~K)
and N$_2$ ($T_{\rm fr} \leq 16-18$~K) \citep{Mumma_Charnley11,2013Sci...341..630Q,Du:2015kz,Oberg:2016fh,Krijt_ea16}.

Despite these promising results, almost nothing is known about the sulfur content in disks.
So far, only CS was detected in dozens of disks and SO was only detected in several younger, actively accreting disks
including the disk of the Herbig A0 star AB~Aur and Class~I sources \citep{2011A&A...535A.104D,2016A&A...592A.124G,PachecoVazquez:2016fc,Sakai:2016fw}.
\citet{2011A&A...535A.104D} could not detect the SO and H$_{2}$S lines with the IRAM 30-m in LkCa15, DM~Tau, GO~Tau and MWC~480.
They have performed a careful analysis of the previously observed CS lines
and the upper limits of the SO column densities and concluded that the best-fit model
requires a C/O ratio of 1.2, yet this model failed to reproduce the H$_{2}$S upper limits by at least one order of magnitude.
The CS emission in TW~Hya reveals a complex radial distribution with a surface density gap at around 95~au
and low turbulent velocities \citep{2016A&A...592A..49T,Teague:2017ch}.
Using the IRAM 30-m antenna, \citet{2012ApJ...756...58C} have observed CCS in DM~Tau and MWC~480 and obtained upper limits on
the CCS column densities in these disks, $\lesssim 1.5\times 10^{12}$~cm$^{-2}$.
Recently, H$_{2}$CS was found in the Herbig Ae disk of MWC~480 by Loomis et al.~(2017; priv. comm.), where CS was also detected.
Finally, \citet{Booth_ea17_SO} have recently detected SO emission in HD~100546 that does not
come solely from the Keplerian disk, but shows an additional non-disk component (a disk wind, a warp, or shocked gas).

This is in contrast to the studies of the ISM, where sulfur-bearing species are commonly observed and used as tracers of
chemical ages \citep[e.g., CCS/NH$_{3}$;][]{Suzuki_ea92}, temperature/shocks \citep[SO, SO$_{2}$, OCS;][]{2014A&A...567A..95E},
X-ray irradiation \citep[SO;][]{Stauber_ea05},
grain processing \citep[SO/H$_{2}$S, SO/SO$_{2}$;][]{Charnley_ea97}, and turbulent transport \citep[CCS/CO;][]{SW2011}.
The S-bearing molecules such as H$_{2}$S, SO, SO$_{2}$, CS, CS$_{2}$, H$_{2}$CS, S polymers, and other
complex S-bearing chains and organics were detected in comets and satellites of Jovian planets in our solar system
\citep[e.g.][]{Bossier_ea07, LeRoy2015, Calmonte_ea16}.
Recently, sulfur-bearing species have been detected in the molecular disk around the late-type Sun-like AGB star \citep{Kervella_ea16}.

The sulfur-bearing amino acids cysteine and methionine as well as OCS acting as a catalyst in peptide synthesis
play an important role in biochemistry and hence sulfur chemistry could be related to the origin of life \citep{Brosnan01062006,Chen_ea15}.
In order to make progress in our understanding of the importance of the delivery of S-bearing molecules to early Earth and Earth-like
planets a better understanding of the sulfur content of protoplanetary disks is required.
Intriguingly, modern astrochemical models cannot fully reproduce the observed abundances or upper limits of sulfur-bearing species
either due to uncertainties in the reaction rates or missing key formation routes or unaccounted major reservoirs of sulfur
\citep[see e.g.][]{Druard:2012bc,Loison:2012fl,Vidal:2017gw}.

The main goals of this paper are twofold. First, we aim to study in detail sulfur chemistry in the DM~Tau
disk using ALMA observations that are several times more sensitive that the previous observations. Second, we aim to investigate
the sensitivity of the disk sulfur chemistry to various physical and chemical parameters, using an appropriate DM~Tau physical-chemical model.
The paper is organized as follows. In Sec.~\ref{sec:obs} our ALMA Cycle~3 observations are presented. Disk-averaged
column densities and upper limits are derived in Section~\ref{sec:obs_nx}.
In Sections~\ref{sec:disk_phys} and \ref{sec:disk_chem}
the adopted physical and chemical models of the DM~Tau disk are described, and the parameter space to be calculated is
presented. The comparison between the observed values and the grid of DM~Tau models is discussed in Sec.~\ref{sec:results}.
Discussion and Conclusions follow.


\section{Observations}
\label{sec:obs}


Observations of DM Tau were carried out with ALMA at June 22 (in Band 6) and July 26, 2016 (in Band 4)
as part of the ALMA project 2015.1.00296.S, PI: Dmitry Semenov.
The phase tracking center was $\alpha_{J2000} = 04\rm^{h} 33\rm^{m}48\fs747$,  $\delta_{J2000} = 18\degr 10\arcmin 09\farcs585$.
The systemic velocity of the source was set to v$_{\rm LSR} = 6$~km\,s$^{-1}$. More specifically, Band~4 data were taken with 45
antennas and baselines from 15.1~m up to 1.1~km with an on source time of about 2~min. The Band 4 spectral setup consisted of six
spectral windows with 480 channels each with a channel width of 122.07~kHz ($\sim 0.3$~km~s$^{-1}$), covering about 0.4~GHz
between 145.926~GHz and 159.010~GHz, and one continuum spectral window with the 2~GHz bandwidth ($128 \times 15.625$~MHz channels)
centered at 145.021~GHz. The quasars J0510+1800 and J0431+2037 were used as bandpass, flux, and phase calibrators.
The Band 6 data were taken with 37 antennas and baselines from 15.1~m up to 704.1~m with an on source time of about 20~min,
using J0510+1800 and J0423-0120 as bandpass, flux, and phase calibrators. The Band 6 spectral setup consisted of three spectral
windows with 240 channels each with a channel width of 244.141~kHz ($\sim 0.3$~km~s$^{-1}$), covering about 0.2~GHz between 215.195~GHz
and 218.936~GHz and one continuum spectral window with the 2~GHz bandwidth ($128 \times 15.625$~MHz channels) centered at 232.012~GHz.

Data reduction and continuum subtraction in the $uv$-space were performed through the Common Astronomy Software Applications (CASA) software
\citep{McMullin_ea07}, version~4.5.3. The continuum image at 145.021~GHz was self-calibrated, and the solutions were applied to the
lines observed in Band~4 in order to improve the signal-to-noise ratio, in particular for the SO$_2$ transitions.
To further optimize the sensitivity, the Bands~4 and 6 data were cleaned using the ``natural'' weighting.
The resulting synthesized beam size is $0.61\arcsec \times 0.58\arcsec$ at a positional angle of $\sim 28\degr$ for the 145~GHz continuum
image, with the $1~\sigma$~rms of $7.6 \times 10^{-5}$~Jy\,beam$^{-1}$. For the line images at the Band~4 and 6 the synthesized beams are similar,
$\approx 0.6\arcsec \times 0.5\arcsec$ (positional angle is $23\degr$), with the spectral resolution of $\sim 0.3$~km\,s$^{-1}$
and the $1~\sigma$~rms sensitivity of $5 \times 10^{-3}$~Jy\,beam$^{-1}$.


\subsection{Line imaging}
\label{sec:obs_line}
\begin{figure}
\centering
\includegraphics[width=0.475\hsize]{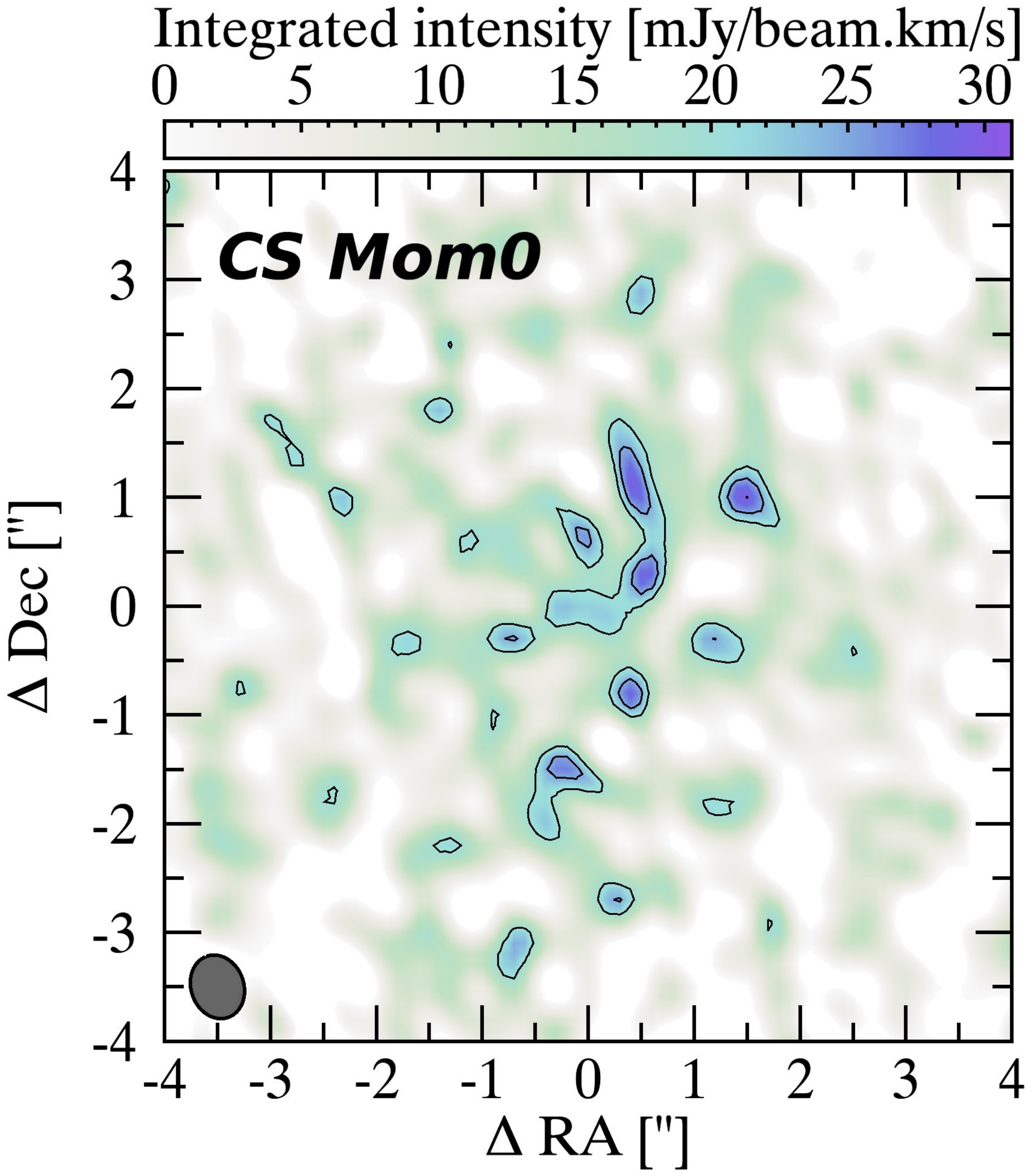}
\includegraphics[width=0.485\hsize]{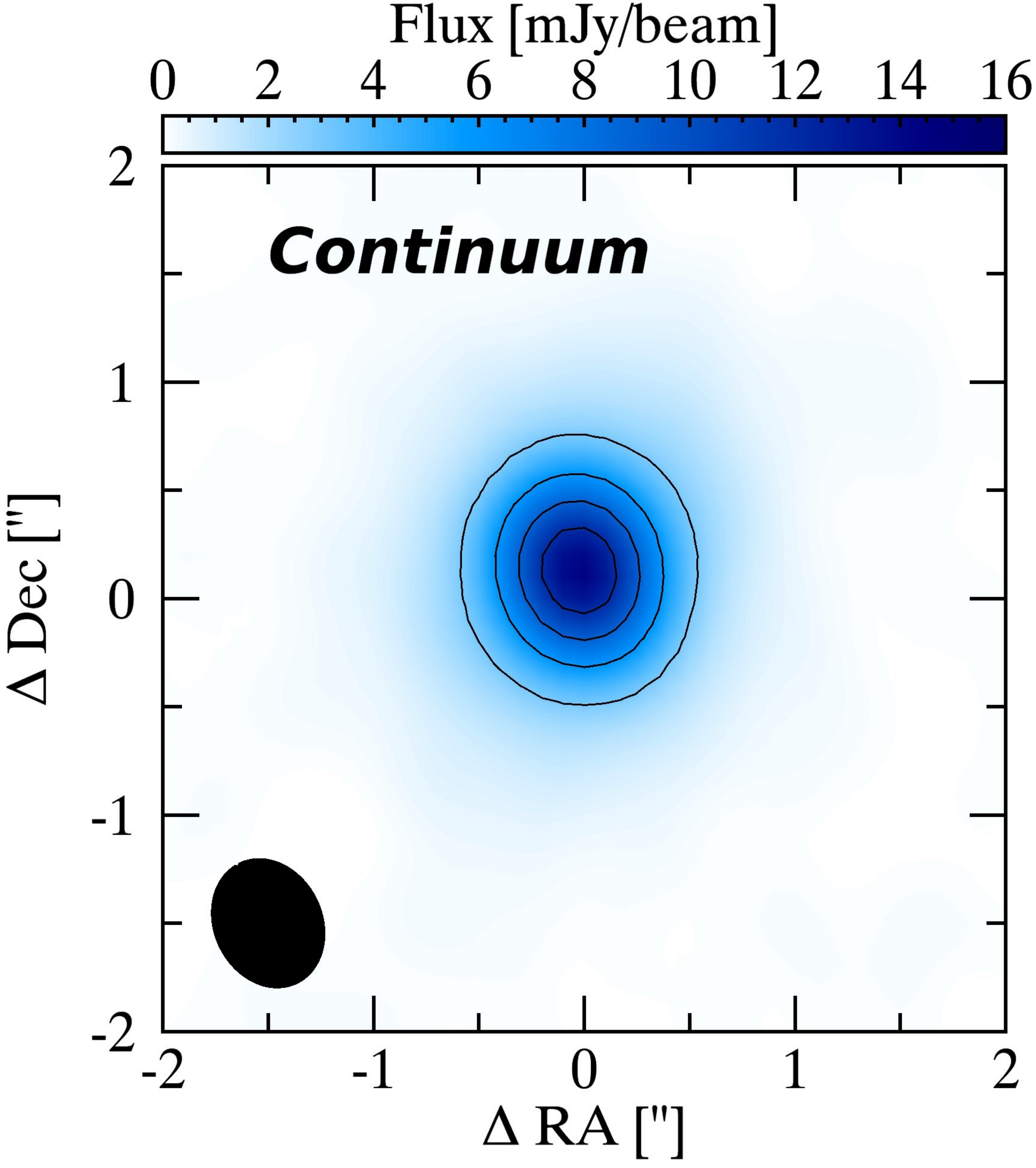}
\caption{\label{fig:CS3-2_0map} (Left) The ALMA Band~4 CS~(3-2) $0$th-moment map of the DM~Tau disk.
(Right) The dust continuum map of the DM~Tau disk at 146.969~GHz. The first contours and the level steps are at $3~\sigma$ rms.}
\end{figure}

\begin{figure}
\centering
\includegraphics[width=0.75\hsize]{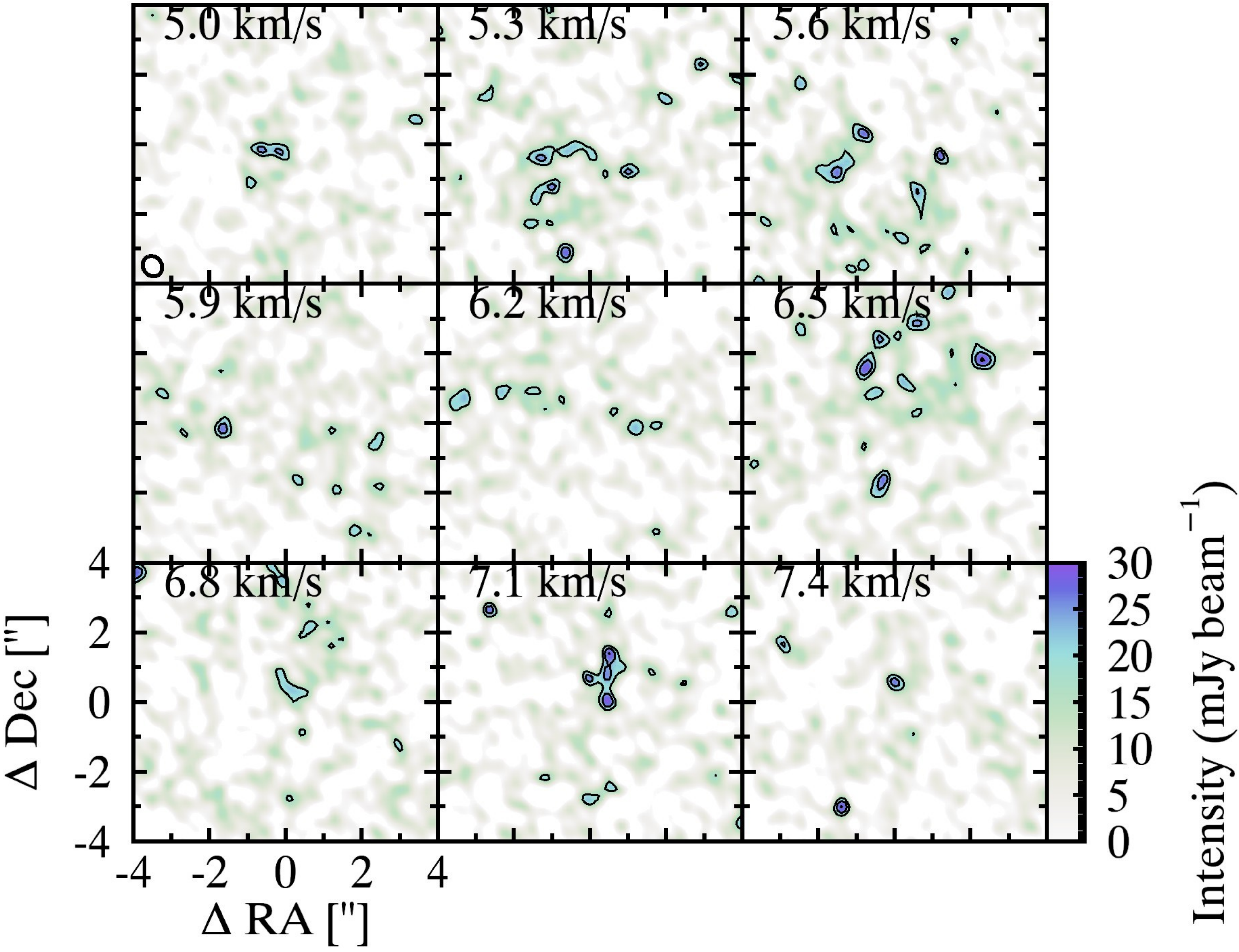}
\caption{\label{fig:CS3-2_map} The ALMA Band~4 spectral map showing CS~(3-2) in the DM~Tau disk.
The channel spacing is $0.3$~km\,s$^{-1}$. The first contour and the level step are at $3~\sigma$ rms.}
\end{figure}

The targeted molecular lines, their frequencies, and spectroscopic parameters are listed in Table~\ref{tablines}.
The only detected emission ($>3\sigma$) visible in the continuum-subtracted integrated intensity and spectral maps was CS~(3-2),
see Figs.~\ref{fig:CS3-2_0map} and \ref{fig:CS3-2_map}.
Both maps are relatively noisy because the ALMA Band~4 integration time was short, only about 2 min on-source.
The observed line parameters assuming a source size of 4$\arcsec$, which is estimated from the CS integrated emission map over the line profile,
are presented in Table~\ref{tabfluxes}.

\begin{figure}
\centering
\includegraphics[width=0.32\hsize]{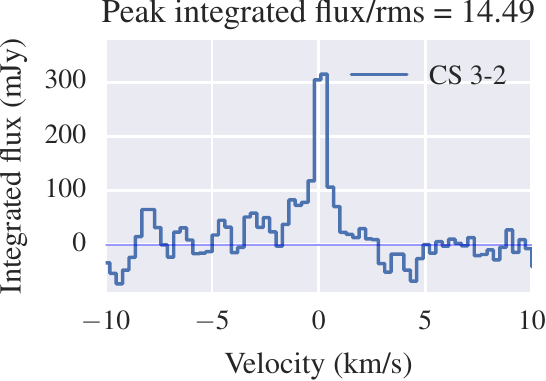}
\includegraphics[width=0.32\hsize]{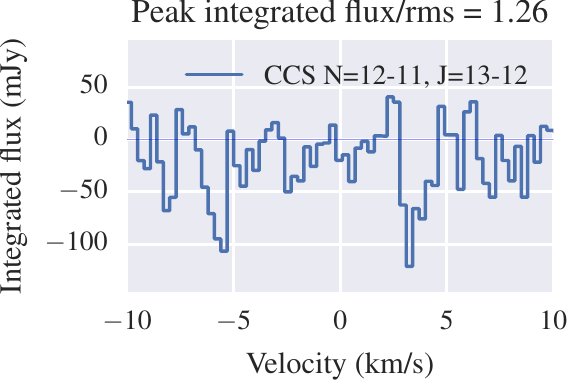}
\includegraphics[width=0.32\hsize]{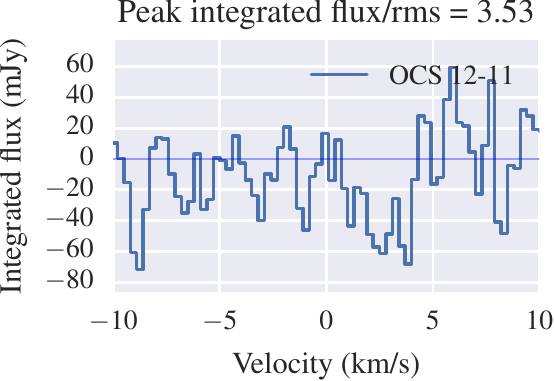}\\
\includegraphics[width=0.32\hsize]{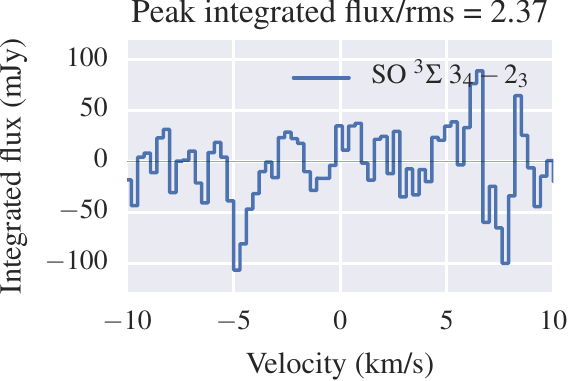}
\includegraphics[width=0.32\hsize]{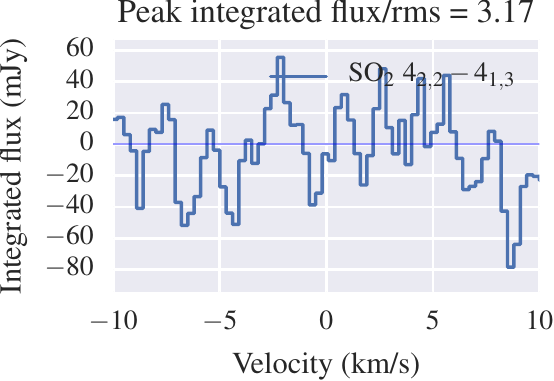}
\includegraphics[width=0.32\hsize]{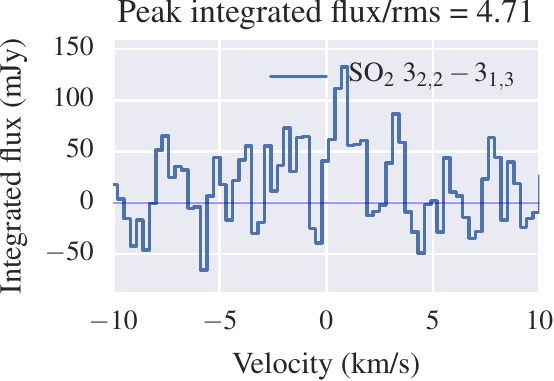}
\caption{\label{fig:spectra_B4} The pixel deprojected and azimuthally averaged ALMA Band~4 spectra showing CS (3-2), CCS (N=12-11, J=13-12),
OCS (12-11), SO ($^{3}\Sigma$ v=0, $3_{4}-2_{3}$), SO$_{2}$ (v=0, $4_{2,2}-4_{1,3}$), and SO$_{2}$ (v=0, $3_{2,2}-3_{1,3}$) spectral windows in DM~Tau.}
\end{figure}

\begin{figure}
\centering
\includegraphics[width=0.32\hsize]{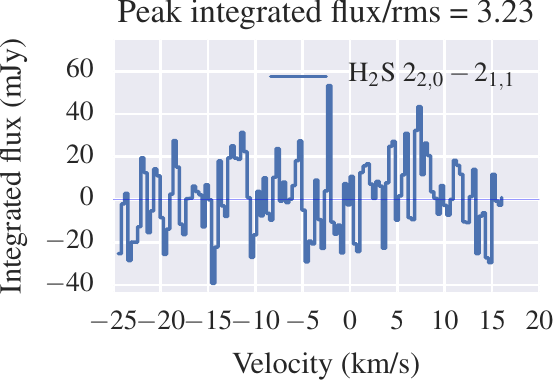}
\includegraphics[width=0.32\hsize]{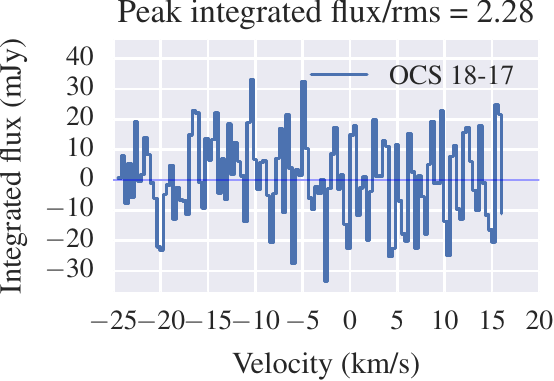}
\includegraphics[width=0.32\hsize]{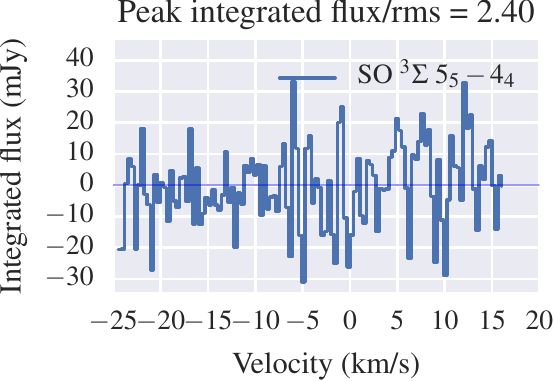}
\caption{\label{fig:spectra_B6} The pixel deprojected and azimuthally averaged ALMA Band~6 spectra showing H$_{2}$S (2$_{2,0}-2_{1,1}$),
OCS (18-17), and SO ($^{3}\Sigma$ v=0, $5_{5}-4_{4}$)  spectral windows in DM~Tau.}
\end{figure}

To increase the signal, we applied the pixel deprojection and azimuthal averaging technique
as presented in \citet{2016A&A...592A..49T}, \citet{Yen_ea_16}, and \citet{Matra_ea17}.
For deprojection we used a inclination angle of $-34\degr$, a positional angle of $27\degr$, a distance of 140~pc,
and a stellar mass of $0.55M_{\sun}$ \citep{Teague:2015jk,Simon_ea17}.
After such deprojection, a double-peaked spectrum typical for Keplerian disks becomes a single-peaked
spectrum with a narrower linewidth.
The resulting shifted spectra are presented in Fig~\ref{fig:spectra_B4} for Band~4 and in Fig.~\ref{fig:spectra_B6}
for Band~6, respectively. As can be clearly seen, the only firmly detected line in our dataset is CS~(3-2), with a peak signal-to-noise ratio
$\gtrsim 14$.
There is a tentative detection of the SO$_{2}$ ($3_{2,2}-3_{1,3}$) line, with a peak signal-to-noise ratio $\approx 4.7$,
albeit the signal is slightly shifted from the velocity center by $+1$~km\,s$^{-1}$ (Fig.~\ref{fig:spectra_B4}, left bottom panel).
The rest of the lines are not detected.

\begin{table*}
\caption{Spectroscopic line parameters\label{tablines}}
\centering
\begin{tabular}{llll}
\hline\hline
Line  & Frequency & $E_{u}$ & $S\mu^{2}$\\
      & (GHz)     & (K)     & ($D^{2}$) \\
\hline
CS 3-2                             & 146.96903 & 14 & 11 \\
CCS N=12-11, J=13-12               & 156.98165 & 51 & 108\\
H$_{2}$S 2$_{2,0}-2_{1,1}$         & 216.71044 & 84 & 2\\   
OCS 12-11                          & 145.94681 & 46 & 9 \\
OCS 18-17                          & 218.90336 & 100 & 9 \\
SO $^{3}\Sigma$ v=0, $3_{4}-2_{3}$ & 158.97181 & 29 & 6 \\
SO $^{3}\Sigma$ v=0, $5_{5}-4_{4}$ & 215.22065 & 44 & 11\\
SO$_{2}$ v=0, $4_{2,2}-4_{1,3}$    & 146.60552 & 19 & 6 \\
SO$_{2}$ v=0, $3_{2,2}-3_{1,3}$    & 158.19974 & 15 & 4 \\
\hline
\end{tabular}
\tablefoot{Col.3 Upper state energies. Col.4 Line strengths.The data are taken from the
CDMS database \citep{Muller2005}, \url{https://www.astro.uni-koeln.de/cdms/catalog}.}
\end{table*}

\begin{table*}
\caption{
\label{tabfluxes}
Line parameters for the observed sulfur-bearing transitions towards DM~Tau.}
\centering
\begin{tabular}{lccclc}
\hline\hline
Molecule & Transition & $v_{\rm LSR}$ & $\Delta v_{\rm FWHM}$ & $I_{\rm peak}$ & $\int$ TdV \\
\hline
 & & (km/s) & (km/s) & (mJy) &  (Jy~km~s$^{-1}$)  \\
\hline
CS & 3-2 & 6.2(0.3)& 1.9(0.6) & 520(100)&1.1(0.3)\\
CCS N=12-11 & 13-12& ... & ...& $\le$130 &... \\
H$_2$S &2$_{2,0}$-2$_{1,1}$& ... & ...& $\le$10 &... \\
OCS v$=$0 & 12-11& ... & ...& $\le$130 &... \\
OCS v$=$0 & 18-17 & ... & ...& $\le$10 &... \\
SO $^{3}\Sigma$ v$=$0 &3$_{4}$-2$_{3}$& ... & ...& $\le$130 &... \\
SO $^{3}\Sigma$ v$=$0& 5$_{5}$-4$_{4}$  & ... & ...& $\le$10 &... \\
SO$_2$ v$=$0 & 4$_{2,2}$-4$_{1,3}$& ... & ...& $\le$350 &... \\
SO$_2$ v$=$0 & 3$_{2,2}$-3$_{1,3}$& ... & ...& $\le$250 &... \\
\hline
\end{tabular}
\tablefoot{The observed values along with the 3$\sigma$ uncertainties (given in brackets) obtained by the Gaussian fit.
Col.3 LSR velocity. Col.4 Full Width Half Maximum line width. Col.5 Peak intensity. Col.6 Line area.}
\end{table*}

\section{Observed CS column density and the upper limits for SO, SO$_{2}$, H$_2$S, OCS and CCS}
\label{sec:obs_nx}

\subsection{Fitting with RADEX}
\label{sec:obs_nx_RADEX}

Since we are mainly dealing with non-detections, and our CS data are noisy, we decided to start with a
simple approach based on \citet{2012A&A...544L...9F,Fedele_ea13a}
to derive disk-averaged vertical column densities or their upper limits. The ALMA data in the CASA format
were exported as FITS files and analyzed in GILDAS\footnote{\url{https://www.iram.fr/IRAMFR/GILDAS/}}.
Each individual integrated spectrum was fitted with a Gaussian function, assuming a source diameter of $4\arcsec$
(560~au at a distance of 140~pc), based on the CS 0th-moment map (see Table~\ref{tabfluxes}). To convert
line fluxes to column densities, we employed a widely used RADEX code\footnote{\url{http://home.strw.leidenuniv.nl/~moldata/radex.html}} \citep{Radex}.
The code uses a uniform physical structure to perform line radiative transfer with LVG method based on statistical equilibrium calculations with
collisional and radiative processes and background radiation. Optical depth effects are accounted for with an escape probability
method. The molecular data were taken from the Leiden Atomic and Molecular Database (LAMDA)\footnote{\url{http://home.strw.leidenuniv.nl/~moldata/}},
see \citet{Schoier2005}. A local line width of 0.14~km~s$^{-1}$ was assumed for all molecules based on the previous CS observations
by \citet{2012A&A...548A..70G}. Note that when 2 transitions were observed, they were both stacked and taken into account.

\begin{figure}
\centering
\includegraphics[width=0.32\hsize]{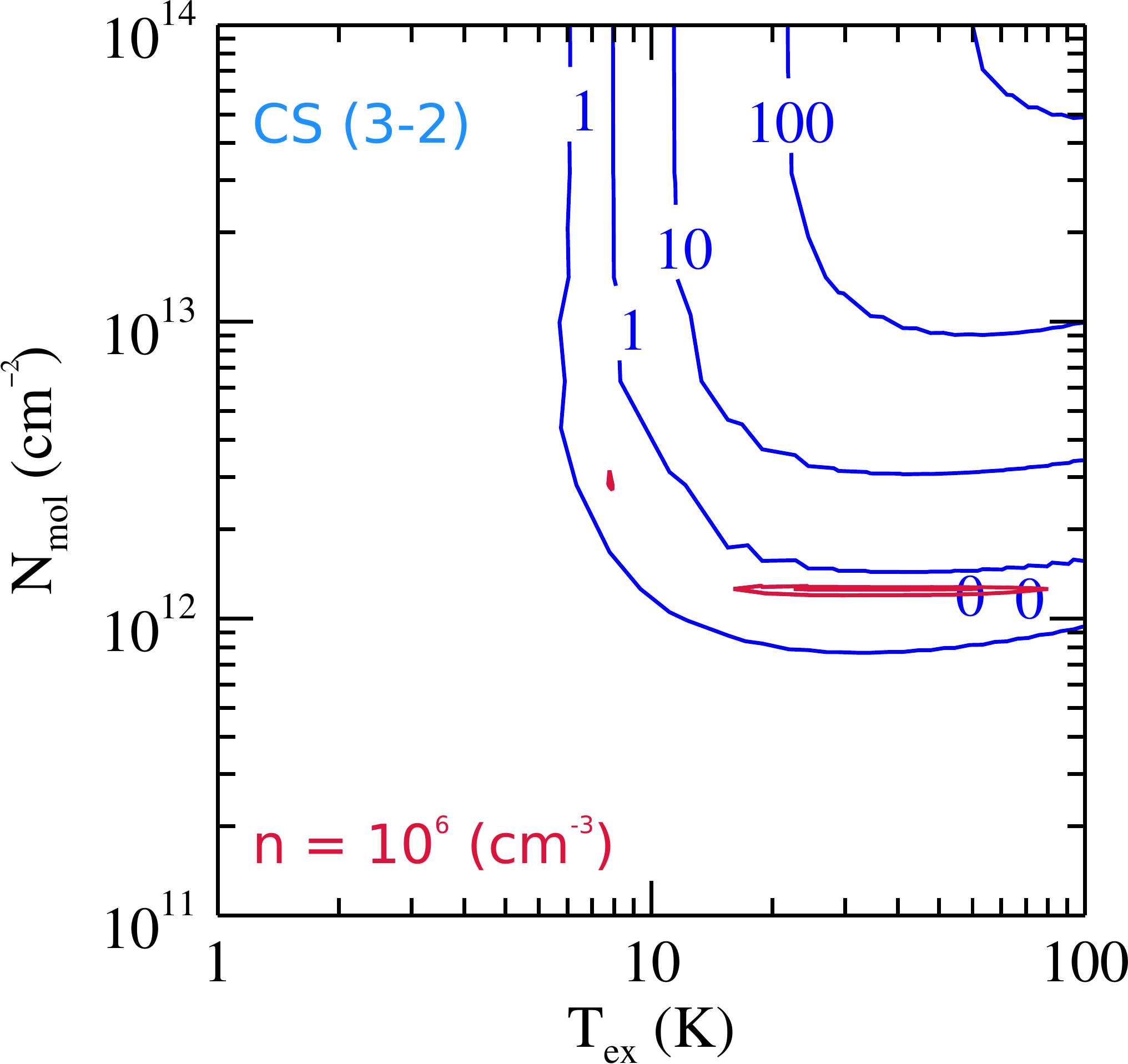}
\includegraphics[width=0.32\hsize]{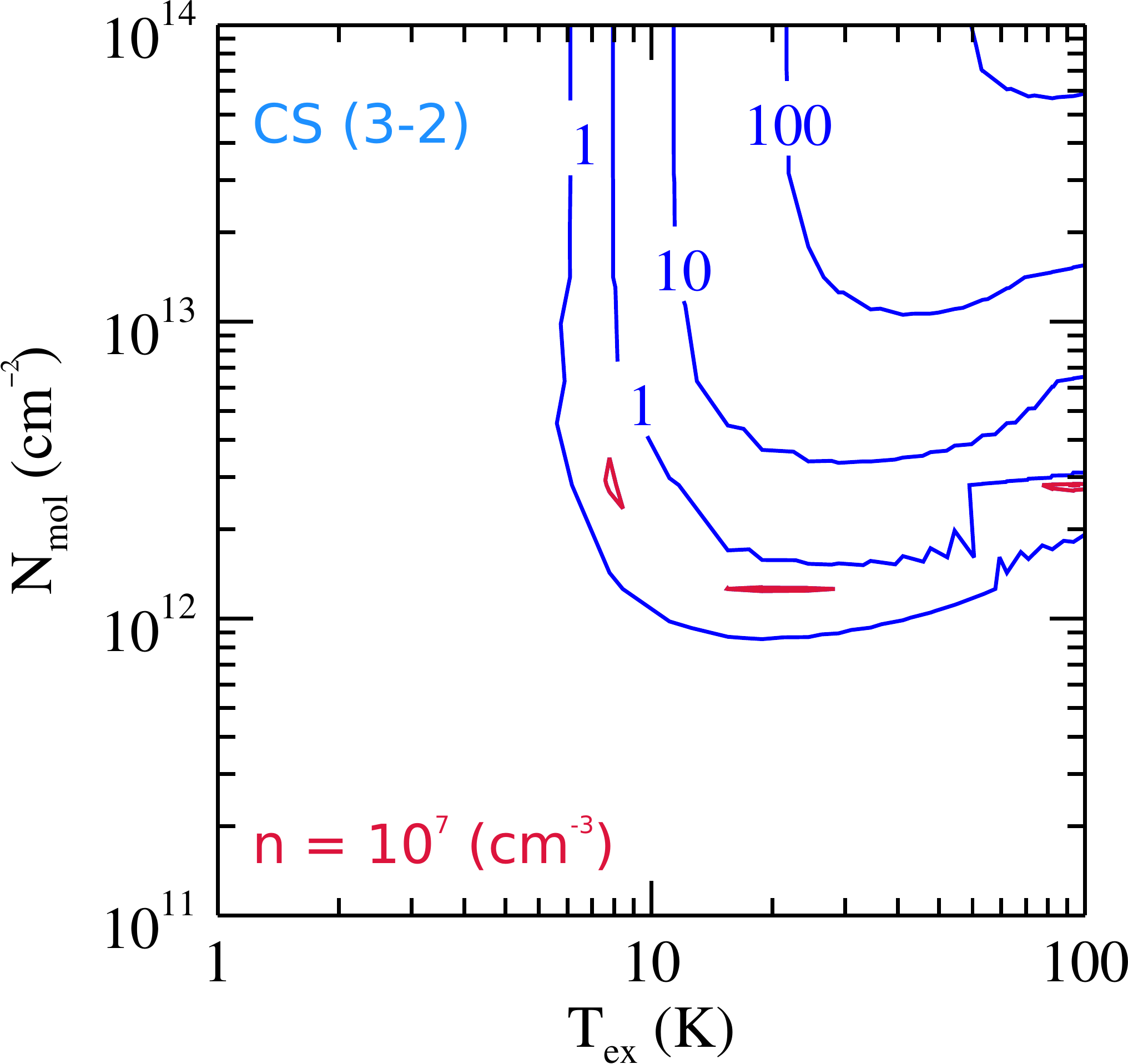}
\includegraphics[width=0.32\hsize]{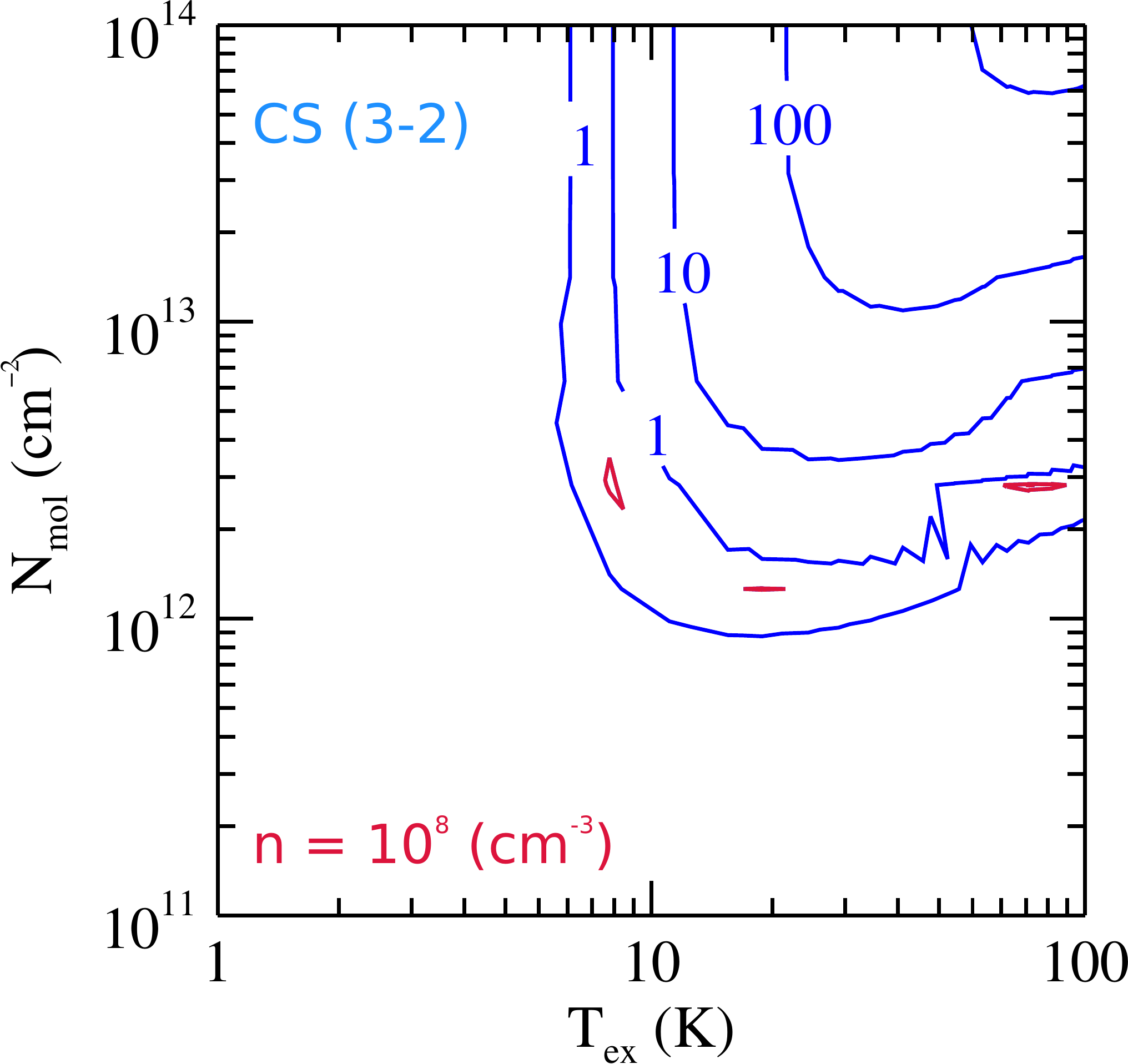}\\
\includegraphics[width=0.32\hsize]{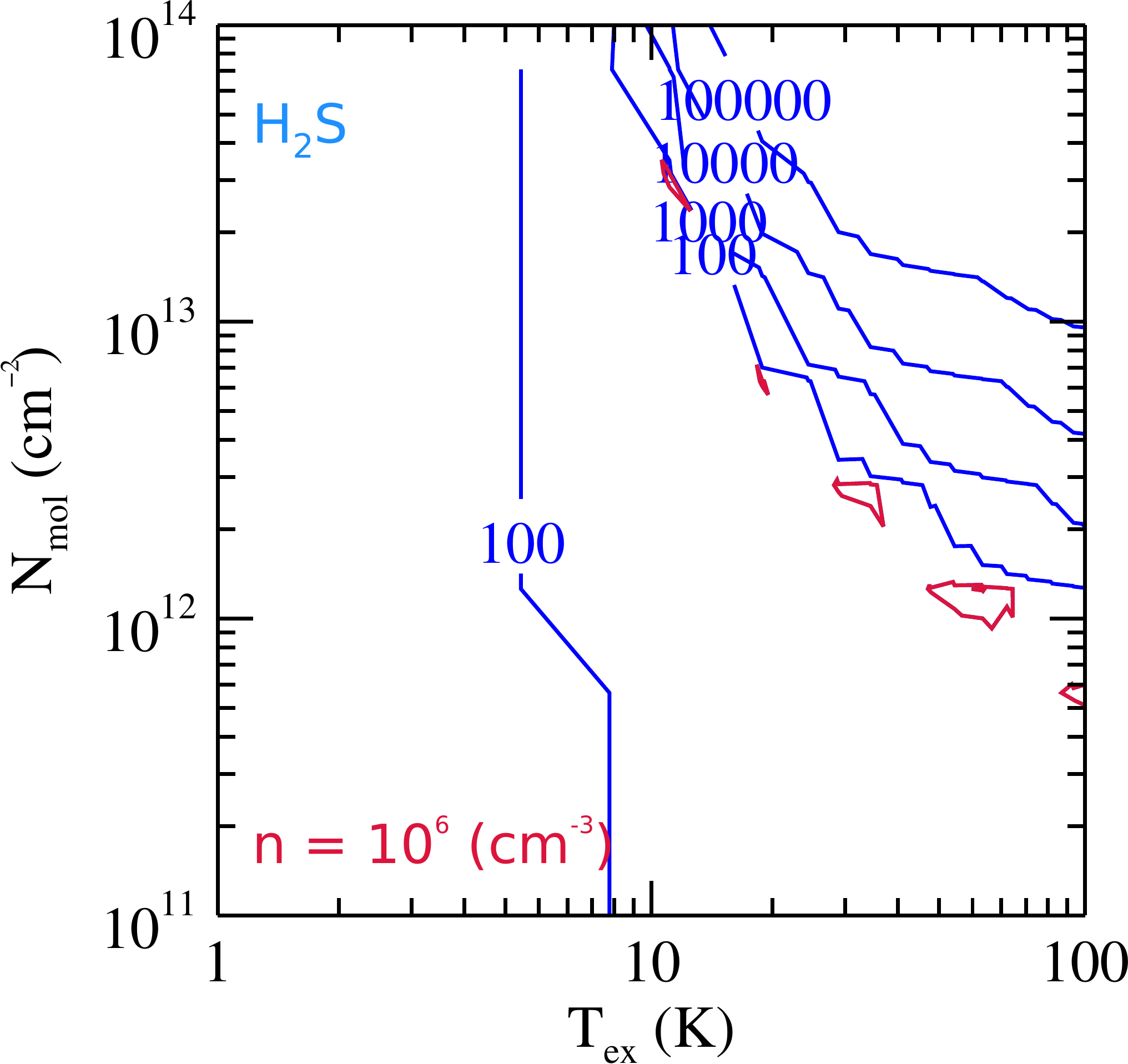}
\includegraphics[width=0.32\hsize]{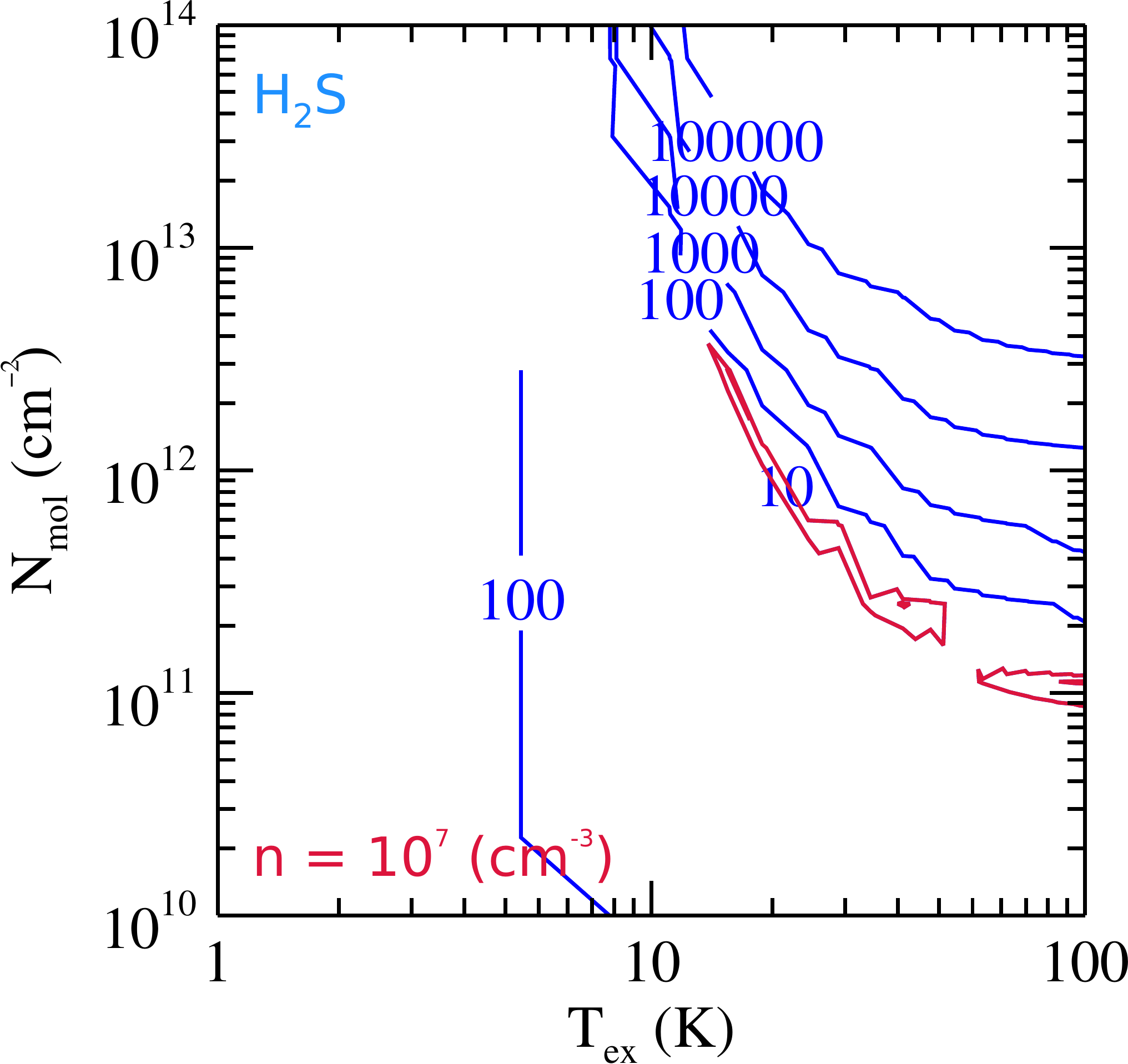}
\includegraphics[width=0.32\hsize]{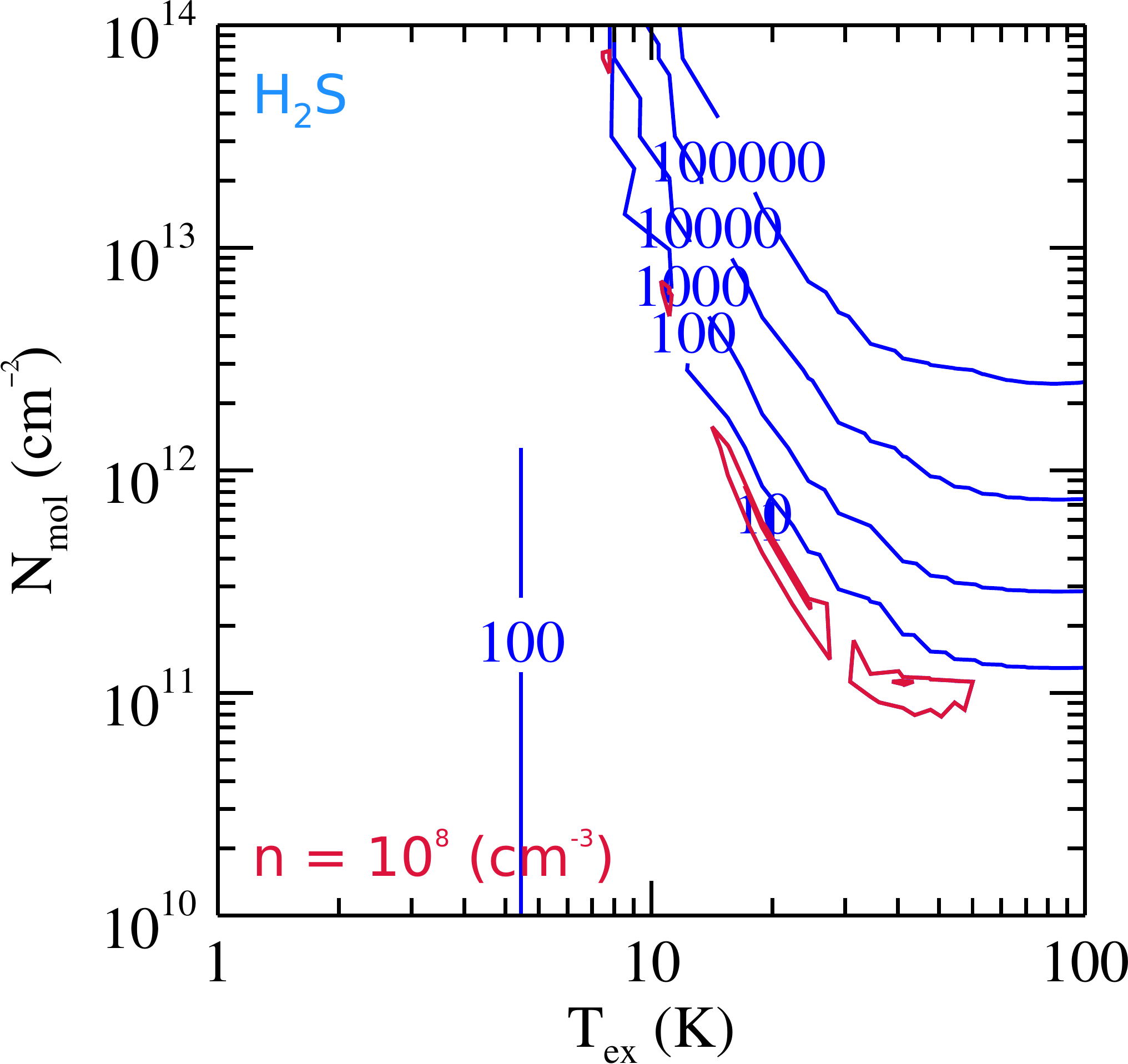}\\
\includegraphics[width=0.32\hsize]{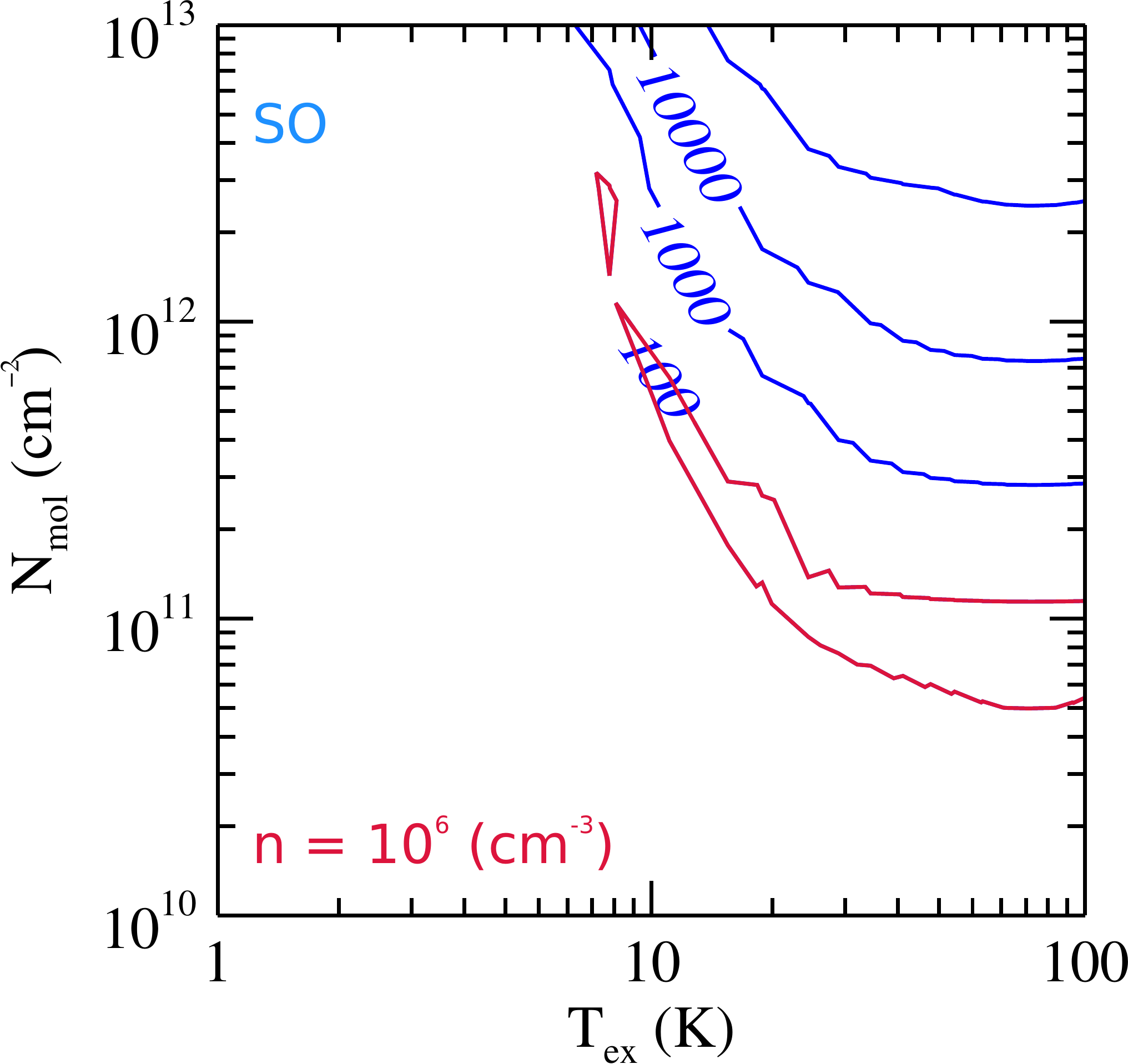}
\includegraphics[width=0.32\hsize]{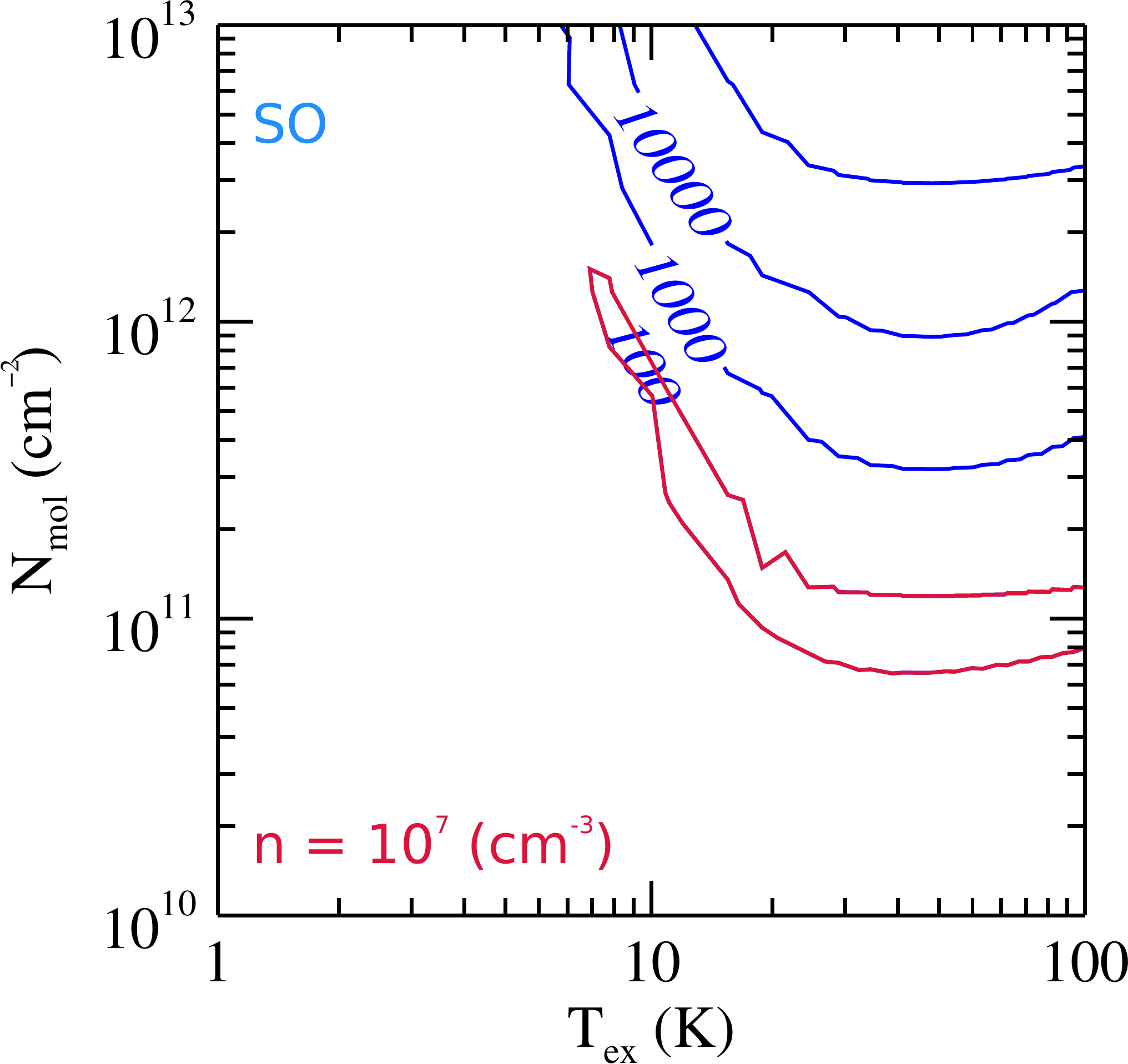}
\includegraphics[width=0.32\hsize]{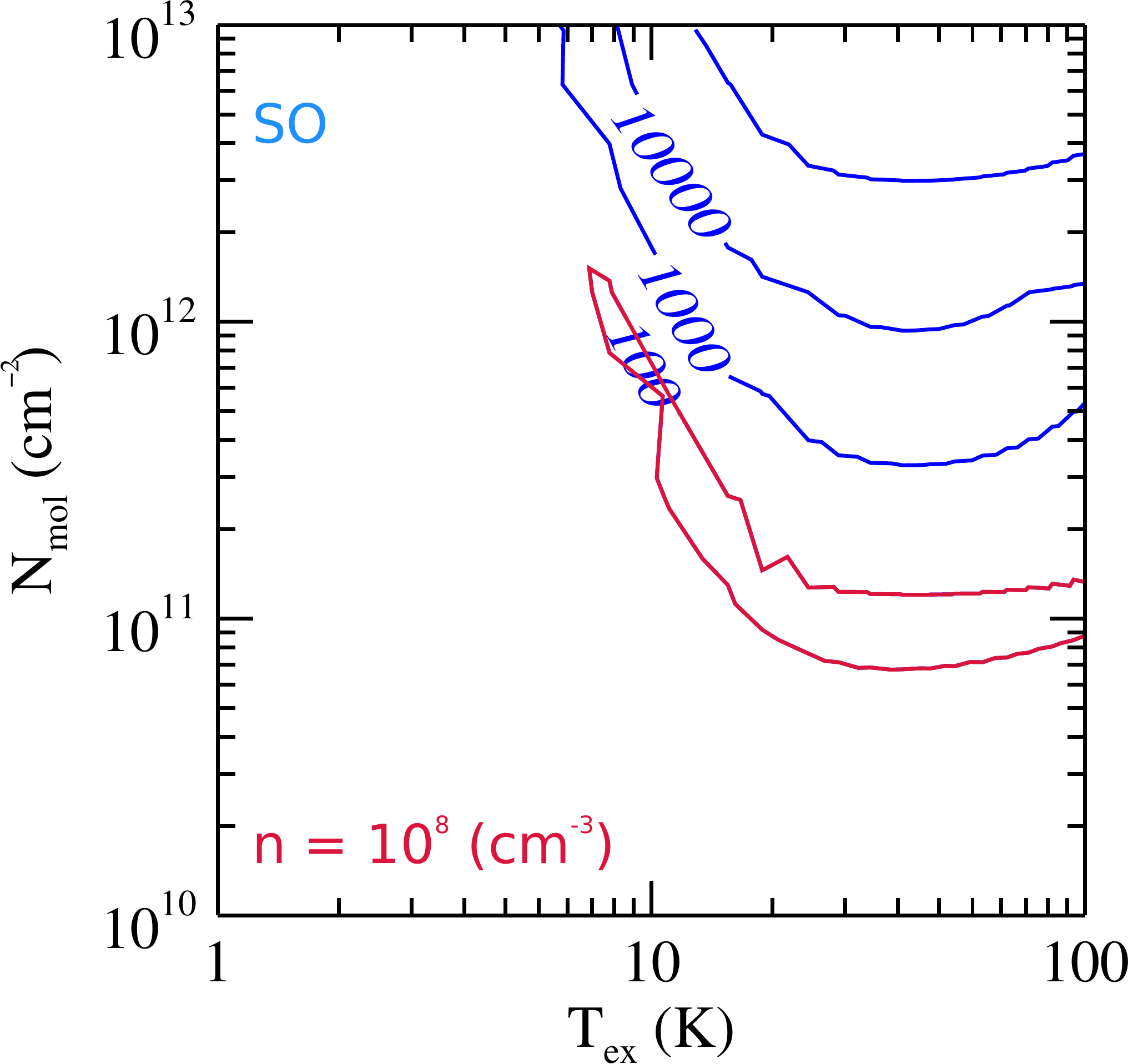}\\
\includegraphics[width=0.32\hsize]{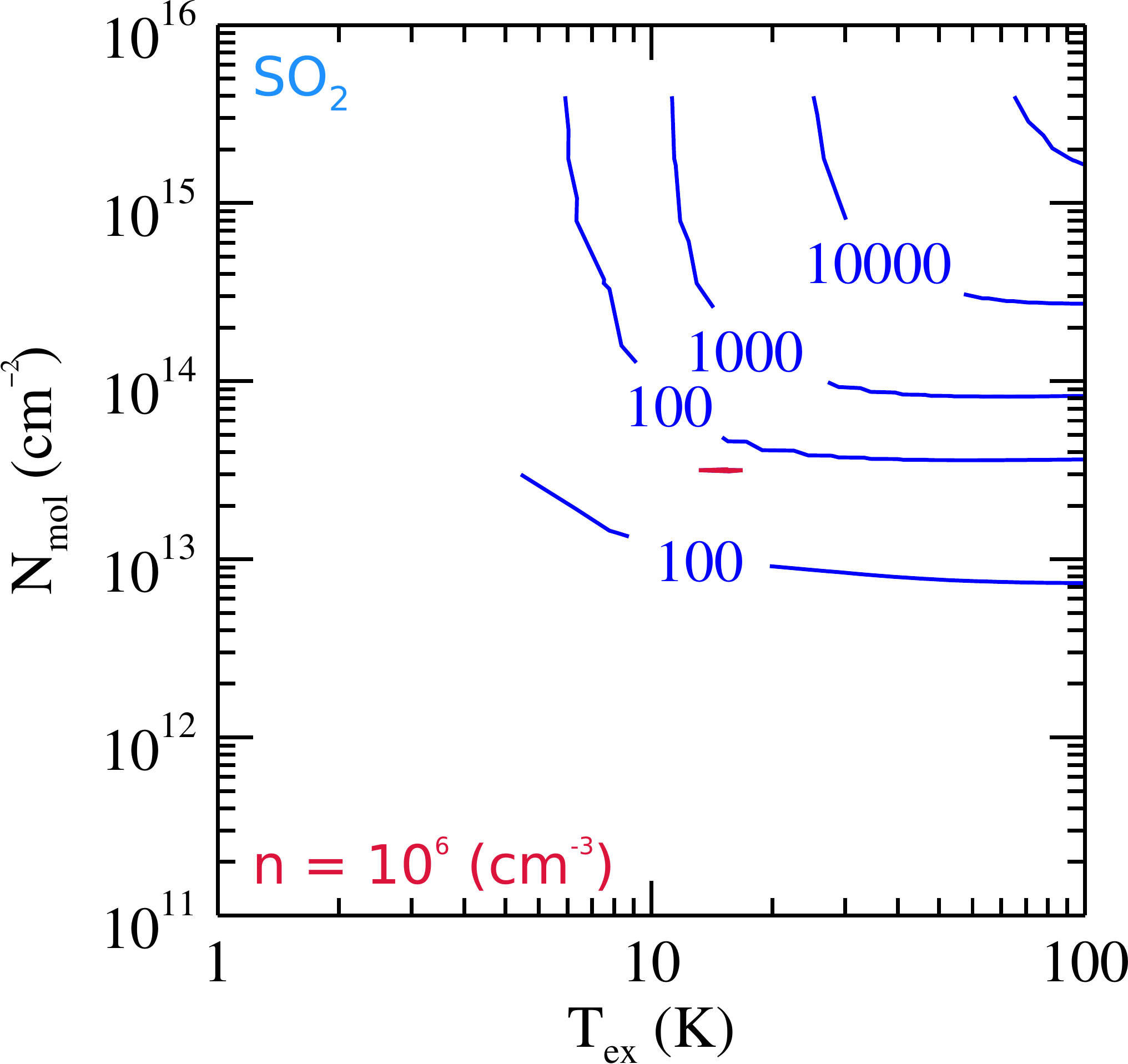}
\includegraphics[width=0.32\hsize]{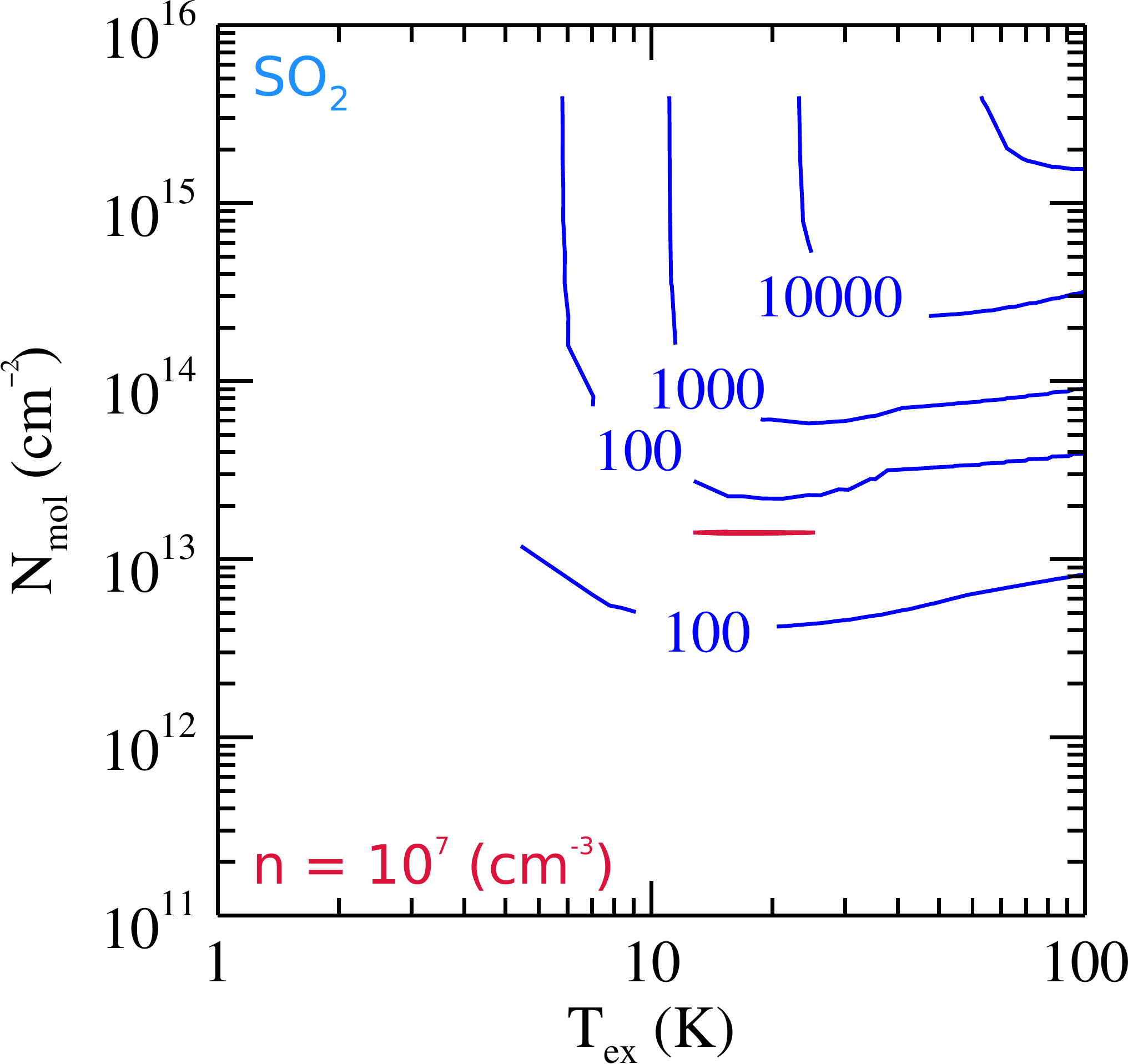}
\includegraphics[width=0.32\hsize]{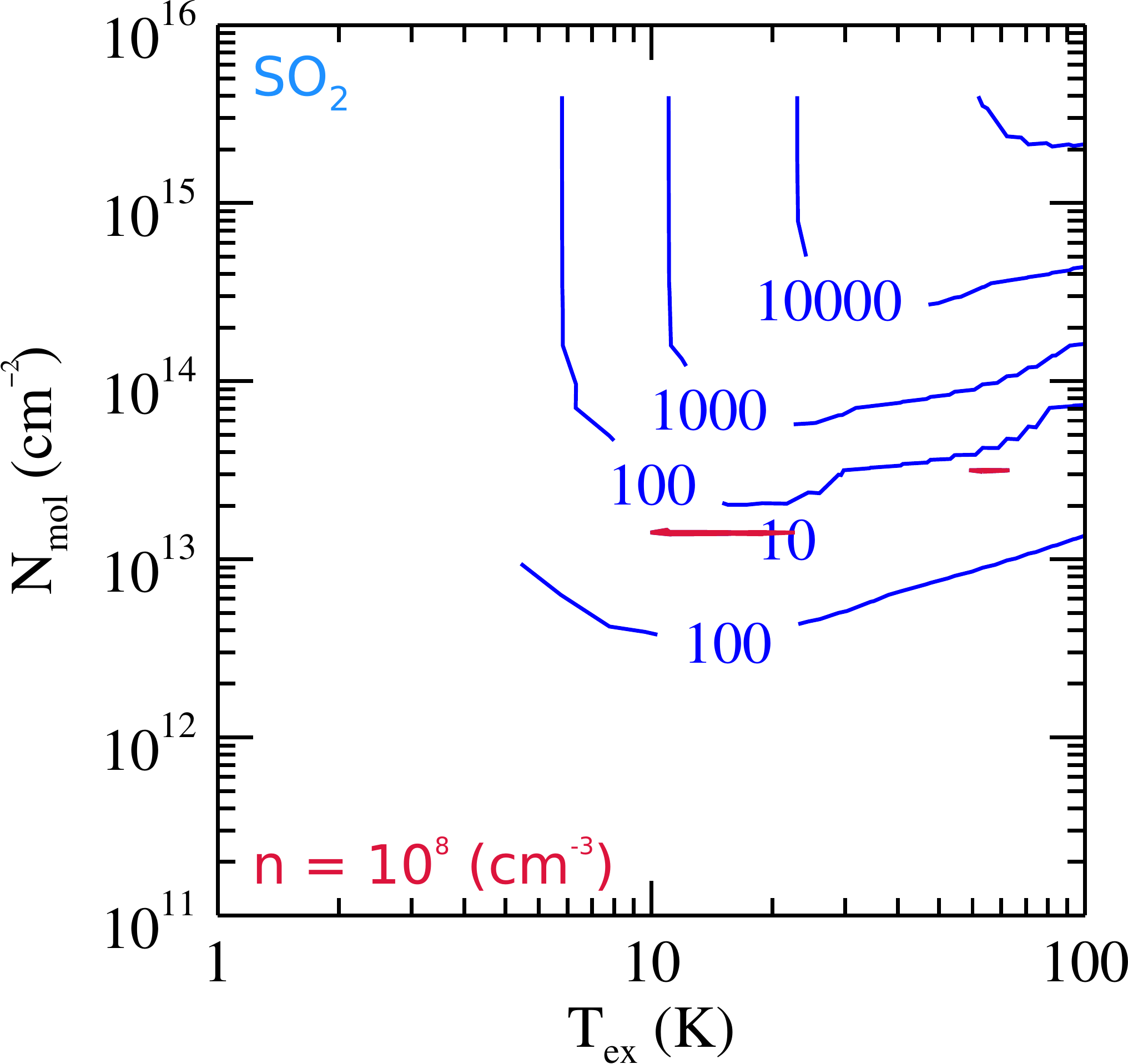}\\
\includegraphics[width=0.32\hsize]{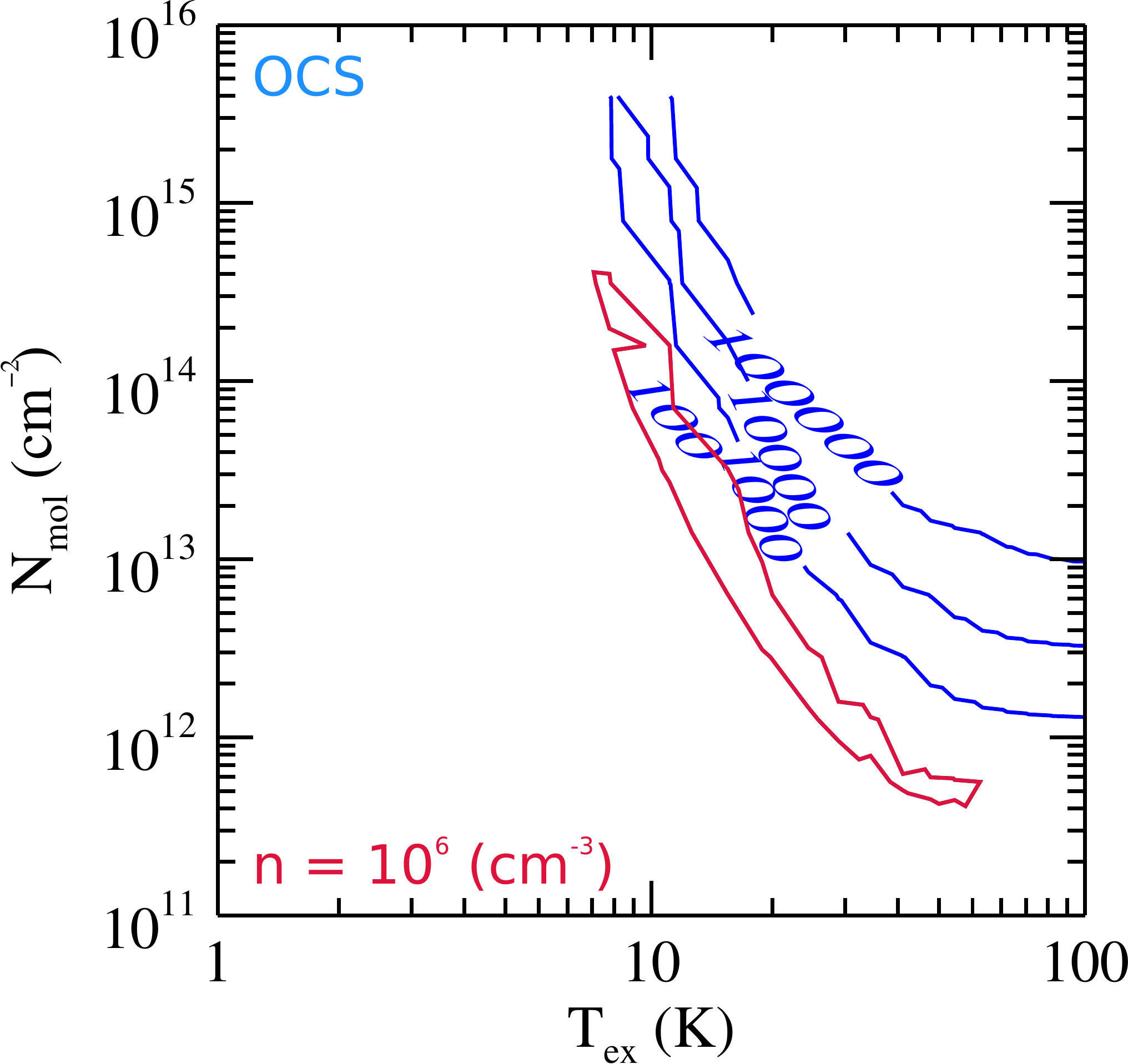}
\includegraphics[width=0.32\hsize]{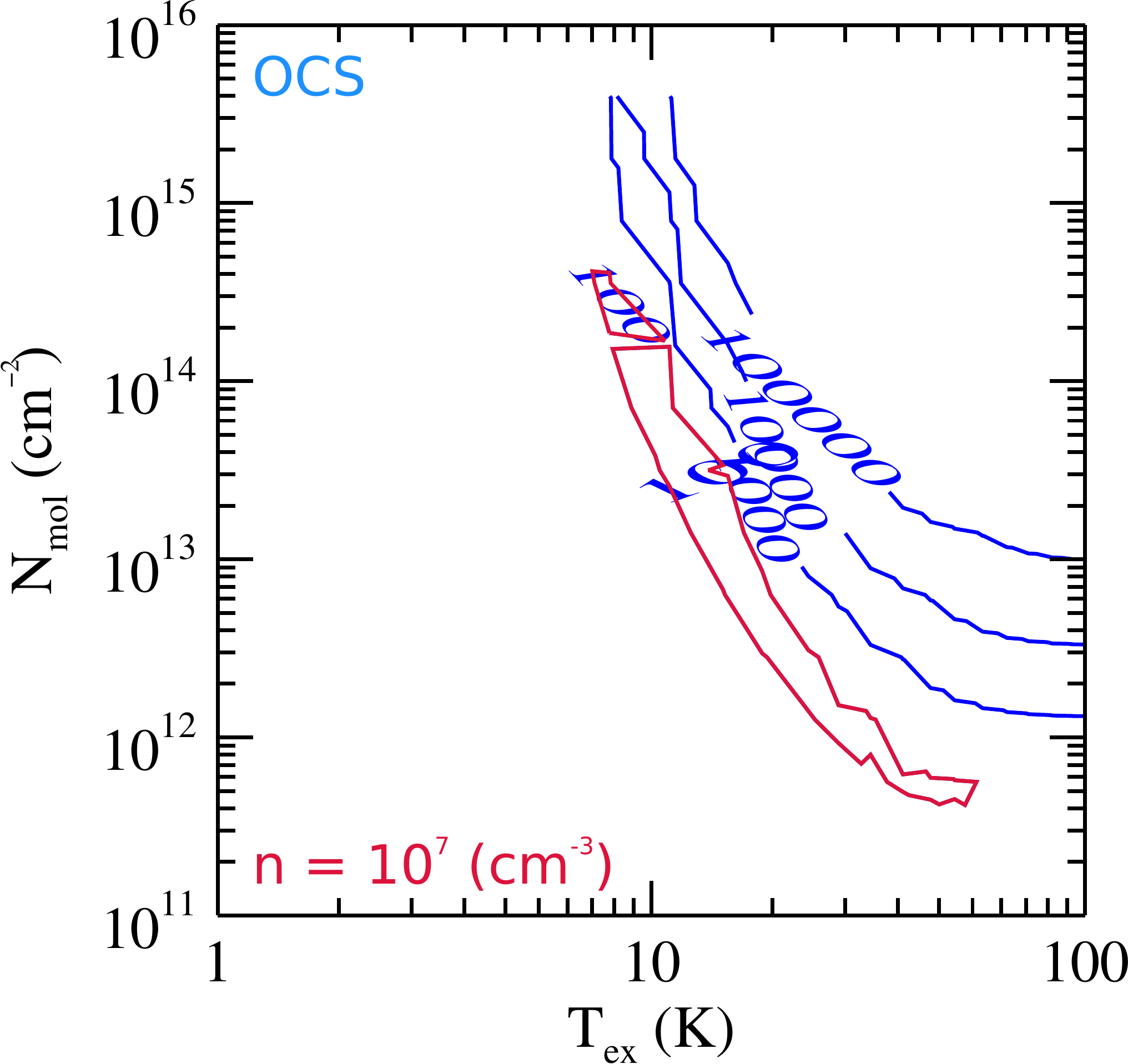}
\includegraphics[width=0.32\hsize]{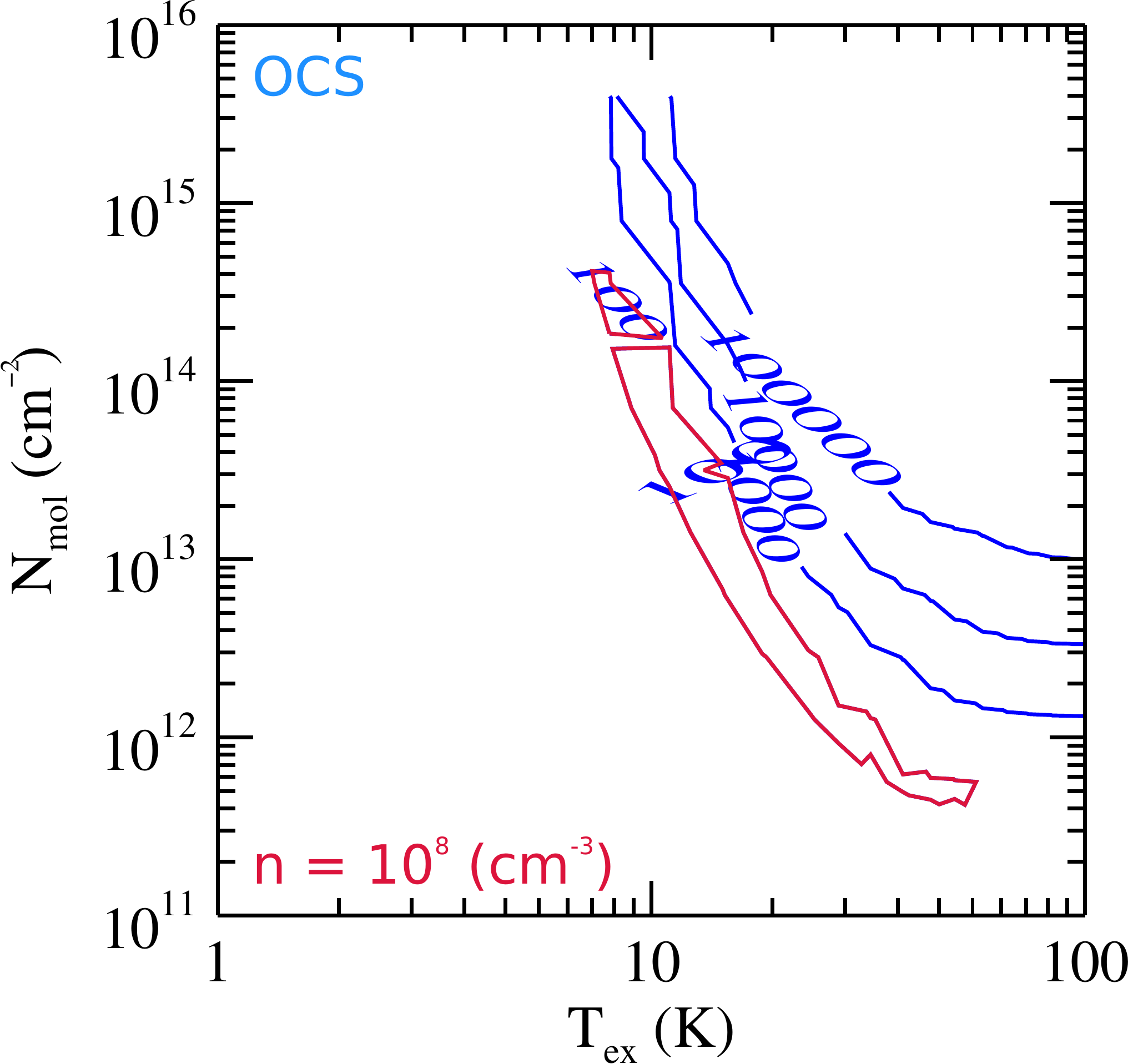}\\
\caption{\label{fig:radex_fit} The $\chi^{2}$ distributions computed with the RADEX code for the three volume densities
of $10^{6}$, $10^{7}$, and $10^{8}$~cm$^{-3}$ and varied $T_{\rm ex}$ and molecular column densities $N$.
(Top to bottom) The results for CS, H$_{2}$S, SO, SO$_{2}$, and OCS are depicted.
The red contours indicate the $\chi^{2}$-minima.}
\end{figure}

By varying excitation temperature, volume density, and column density, the $\chi^{2}$ values between the observed and modeled CS line fluxes
were computed and minimized. The upper limits of the column densities of the other non-detected sulfur-bearing species
were derived in a similar manner, using the observed $3\sigma$ rms as the line flux limit. We were not able to analyze the
CCS data as there are no collisional rate data in the LAMDA and Basecol\footnote{\url{http://basecol.obspm.fr/}} databases.
The resulting $\chi^{2}$ distributions for the assumed volume densities of $10^{6}$, $10^{7}$, and $10^{8}$~cm$^{-3}$
representative of the outer DM~Tau molecular layer are shown in Fig.~\ref{fig:radex_fit}.

The computed $\chi^{2}$ distributions do not depend on the assumed gas density since we have chosen
the most easily excited and strong transitions for our observations. The only exception is H$_{2}$S (2$_{2,0}-2_{1,1}$)
that is slightly better fitted with the high volume density of $10^{8}$~cm$^{-2}$.
Due to their low upper state energies $E_{u} < 20$~K (Table~\ref{tablines}), the temperature dependence of the CS and SO$_{2}$ fits is rather flat,
with the best-fit values of $T_{\rm ex} \lesssim 100$~K.
The higher upper state energies of the H$_{2}$S (2$_{2,0}-2_{1,1}$), OCS (18-17), and SO ($^{3}\Sigma$ v=0, $5_{5}-4_{4}$)
are 84, 100, and 44~K, respectively, which makes them sensitive to the temperature and leads to
a prominent temperature dependence of their $\chi^{2}$ distributions (Fig.~\ref{fig:radex_fit}).
The best-fit temperatures and column densities (or their upper limits) are summarized in Table~\ref{tablinefit}.

\subsection{Parametric fitting with DiskFit}
\label{sec:obs_nx_DiskFit}

As the next step, we performed a more elaborate analysis of the data using our DiskFit parametric method
\citep[see][]{Pietu_ea07,2011A&A...535A.104D,Teague:2015jk}.
The disk structure is parameterized by radial power laws for the surface density, temperature, kinematics, and
molecular column density, which are fitted with a combination of $\chi^2$ minimization and MCMC
of the observed visibilities in the $uv$-plane.
DiskFit first finds the systemic velocity, the geometrical parameters of the disk system (position of the center, inclination and positional angles,
radius of emission), and the disk rotation profile. Next, the dust continuum emission is fitted and dust parameters are obtained.
Finally, gas surface density, scale height, temperature, micro-turbulent velocity,
and column density distribution are derived.

To increase the signal-to-noise ratio of the CS spectrum, the ALMA data were merged with our previous CS~(3-2) IRAM 30-m data
\citep[][]{2011A&A...535A.104D}. Neglecting the primary beam correction, which is justified for our
$\sim 4\arcsec$ emission source, the two visibility sets were
merged after resampling in velocity to the coarsest spectral resolution. For the analysis in DiskFit, we used both datasets
at their native spectral resolution and fitted them assuming the same underlying physical model.
While this approach does not provide a better line image,
it preserves the full information on the line widths at the highest possible accuracy.
To analyze the non-detected lines we used the best-fit temperature and the radial profiles from the merged CS data.
Finally, to get more accurate upper limits for the column densities of SO, SO$_{2}$ and OCS,
we stacked each of their pairs of observed transitions together (see Figs.~\ref{fig:spectra_B4}--\ref{fig:spectra_B6}).
The DiskFit results at a representative outer disk radius of 300~au are given in Table~\ref{tablinefit}.

\begin{table*}
\caption{The best-fit temperatures and column densities or the $3\sigma$ upper limits obtained with RADEX and DiskFit.
\label{tablinefit}}
\centering
\begin{tabular}{|l|l|l|l|l|l|l|l|}
\hline\hline
& \multicolumn{2}{c|}{RADEX}                        & \multicolumn{4}{c|}{DiskFit, representative radius $r=300$~au}\\
\cline{2-7}
Molecule & $T_{\rm ex}$ & $N(X)$                             & $T_{\rm ex}$    & Exponent  & $N(X)$       & Exponent\\
         & (K)          & (cm$^{-2}$)                        & (K)             & $q$         & (cm$^{-2}$) &  $p$   \\
\hline
CS       & $\lesssim 100$    & $1.5-3\times10^{12}$               &  $10\pm2$               & $0.5\pm0.2$  & $6\pm3^{a} \times10^{12}$ & $0.7\pm0.3$   \\
H$_{2}$S & $\lesssim 100$    & $\lesssim 10^{11}-3\times10^{13}$        &  $\textit{10}^{b}$ & $\textit{0.5}$       & $\lesssim 3\times10^{13}$  & $\textit{0.7}$  \\
OCS      & $\lesssim 100$     & $\lesssim 5\times10^{11}-4\times10^{14}$ &  $\textit{10}$           & $\textit{0.5}$ & $\lesssim8 \times10^{11}$ & $\textit{0.7}$   \\
SO       & $\lesssim 100$    & $\lesssim 7\times10^{10}-3\times10^{12}$ &  $\textit{10}$    & $\textit{0.5}$ & $\lesssim4 \times10^{12}$ & $\textit{0.7}$    \\
SO$_{2}$ & $\lesssim 100$    & $\lesssim 1-3\times10^{13}$       &  $\textit{10}$           & $\textit{0.5}$ & $\lesssim10^{13}$         & $\textit{0.7}$   \\
CCS      & ...                  & ...                                &  $\textit{10}$           & $\textit{0.5}$ & $\lesssim10^{14}$         & $\textit{0.7}$   \\
\hline
\end{tabular}
\tablefoot{$^{a}$ The CS is detected at a more than $10\sigma$ signal level, but the derived column density error reflects
the uncertain partition function. $^{b}$ Please note that for the DiskFit analysis of the sulfur-bearing species other than CS the best-fit CS temperature
and the same type of the radial profile were assumed.}
\end{table*}

As can be clearly seen, while the CS column densities derived by both methods are comparable, $\sim 2-6\,10^{12}$~cm$^{-2}$,
the RADEX-based approach cannot constrain CS excitation temperature, as only one transition of CS was observed.
More advanced DiskFit approach is able to recover radial gradients and points to rather cold CS excitation temperatures of $\approx 10$~K
in the outer DM~Tau disk at $\sim 300$~au. This is consistent with the low, $\sim 7-15$~K excitation temperatures for CO, CCH, CN, HCN, CS estimated by
\citet{Pietu_ea07,Henning_ea10,2012A&A...537A..60C,2012A&A...548A..70G}.
The further differences between the methods are mainly due to a) differences in the underlying modeling assumptions,
where RADEX fits the disk-averaged quantities while
DiskFit recovers radial gradients, and b) a left-over degeneracy between the inferred temperature and column density, where
both ``cold/high column density'' and ``warm/low column density'' solutions could exist.
Note that for further theoretical analysis of these results in Section~\ref{sec:results},
it is essential that the both RADEX and DiskFit converge to a similar
best-fit column density ratio for CS/SO of $\gtrsim 1-3$.

\section{Modeling}
\label{sec:model}

\subsection{Disk physical structure}
\label{sec:disk_phys}
\begin{figure}
\centering
\includegraphics[width=0.47\columnwidth,angle=90]{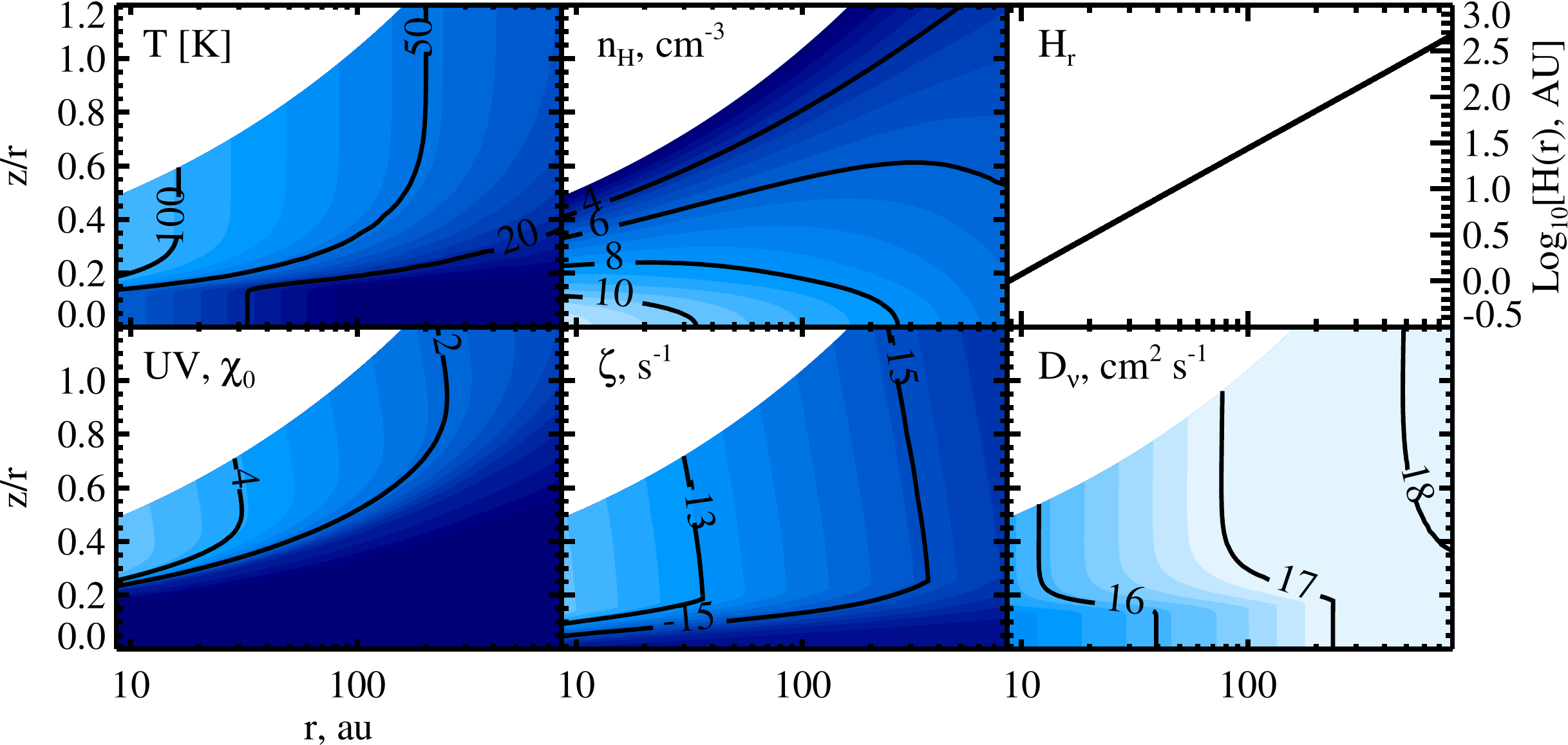}
\caption{\label{fig:dstruc}
The adopted DM~Tau disk physical structure. (Top row, left to right) The spatial distributions of the gas and dust temperature (Kelvin),
the gas density (cm$^{-3}$; $\log_{10}$ scale), and the pressure scale height (au; $\log_{10}$ scale).
(Bottom row, left to right) The UV intensity (in $\chi_{0}$ units of the \citet{G} interstellar FUV field),
the combined ionization rate due to CRPs, stellar X-rays, and short-lived radionuclides (s$^{-1}$), and the
turbulent diffusion coefficient (cm$^2$\,s$^{-1}$; $\log_{10}$ scale).}
\end{figure}


The DM~Tau physical disk model is based on a 1+1D steady-state $\alpha$-model similar to that
of \citet{DAea99}, where equal gas and dust temperatures are assumed. This flaring model was extensively
used in our previous studies of DM~Tau-like disk chemistry
\citep[e.g.,][]{Henning_ea10,SW2011}. The assumption of equal dust and gas temperatures
is reasonably accurate for the molecular layers of S-bearing species that do not extend very high
into the disk atmosphere \citep[see, e.g.][]{ANDES,Woitke_ea09}.
The disk model has an outer radius of $800$~au, an accretion rate of $5\times10^{-9}\mathrm{M}_\odot$\,yr$^{-1}$, a viscosity
parameter $\alpha = 0.01$, and a total gas mass of $0.066\mathrm{M}_\odot$
\citep{Dutrey_ea07,Henning_ea10,SW2011}. This value is close to the upper value of the DM~Tau disk mass, $4.7\times10^{-2}\mathrm{M}_\odot$,
inferred from the HD {\it Herschel} observations by \citet{McClure_ea16}.

The Shakura-Sunyaev parametrization \citep{1973A&A....24..337S} of the
turbulent viscosity $\nu$ was adopted:
\begin{equation}
 \nu(r,z) = \alpha\,c_{\rm s}(r,z)\,H(r),
 \end{equation}
where $H(r)$ is the disk pressure scale height, $c_{\rm s}$ is the sound speed, and
$\alpha = 0.01$ is the dimensionless parameter.

To model turbulent transport along with disk chemical evolution, we used the standard
description of the diffusivity coefficient:
\begin{equation}
D_{\rm turb}(r,z) = \nu(r,z)/Sc,
\end{equation}
where $Sc$ is the Schmidt number that describes the efficiency of turbulent diffusivity
\citep[see e.g.][]{2004ApJ...614..960S,SW2011}.
We considered  three regimes of mixing, namely, laminar disk with no mixing ($Sc=\infty$),
``slow'' mixing ($Sc=100$), and ``fast'' mixing
($Sc=1$). Diffusion of ices is treated similarly to the gas-phase molecules.

The calculated thermal and density structure of the disk is shown in Fig.~\ref{fig:dstruc}.

\subsection{Disk chemical model}
\label{sec:disk_chem}
The adopted chemical model is based on the public gas-grain \texttt{ALCHEMIC} code\footnote{\url{http://www.mpia.de/homes/semenov/disk_chemistry_OSU08ggs_UV.zip}}
\citep[see][]{Semenov_ea10,SW2011}. The chemical network is based on the osu.2007 ratefile with recent updates to
the reaction rates from the Kinetic Database for Astrochemistry (KIDA) \citep{KIDA}.

To calculate UV ionization and dissociation rates, the mean FUV intensity at
a given disk location is obtained by summing up the stellar UV flux $\chi_*(r)=410\chi_{0} \times (r, au/100)^{-2}$
and interstellar UV flux $\chi_{\rm IS}=1-100 \chi_{0}$, which are scaled down by the visual extinction in the radial and
vertical directions, respectively. We used the interstellar UV radiation field $\chi_{0}$ of \citet{G}.
Several tens of photoreaction rates are adopted from
\citet{vDea_06}~\footnote{\url{http://www.strw.leidenuniv.nl/~ewine/photo}}. The self-shielding of
H$_2$ from photodissociation is calculated by  Eq.~(37) from \citet{DB96}. The shielding of CO by
dust grains, H$_2$, and the CO self-shielding is calculated using a precomputed table of
\citet[][Table~11]{1996A&A...311..690L}.

The stellar X-ray radiation is modeled using Eq.~(4; 7-9) from \citet{zetaxa,1997ApJ...485..920G}, with an exponent $n = 2.81$, a cross section at
1~keV of $\sigma_{-22} = 0.85 \times10^{-22}$~cm$^2$, and total X-ray luminosity of $L_{\rm XR} = 10^{31}$~erg~s$^{-1}$.
The X-ray emitting source is assumed to be 12 stellar radii above the star. This makes X-ray ionization to dominate over the CRP ionization
in the disk molecular layer.
The standard cosmic ray (CR) ionization rate is assumed to be $\zeta_{\rm CR}=1.3\times10^{-17}$~s$^{-1}$.
Ionization due to the decay of short-living radionuclides is taken into
account, with the SLR ionization rate of $\zeta_{\rm RN}=6.5\times10^{-19}$~s$^{-1}$ \citep{FG97}.

The gas-grain interactions include sticking of
neutral species and electrons to dust grains with 100\% probability and desorption of ices by
thermal, CRP-, and UV-driven processes.
Uniform amorphous
silicate particles of olivine stoichiometry were used, with a density of $3$~g\,cm$^{-3}$ and a radius of
$0.1\,\mu$m. Each grain provides $1.88\times10^6$ surface sites
for surface recombinations \citep[][]{Bihamea01}, which proceed through the Langmuir-Hinshelwood mechanism
\citep[e.g.][]{HHL92}. The UV photodesorption yield of $10^{-5}$ was adopted \citep{Cruz_Diaz2016,Bertin2016}.
Photodissociation processes of solid species are taken from \citet{Garrod_Herbst06,SW2011}.
We assumed a 1\% probability for chemical desorption
\citep{2007A&A...467.1103G,2013ApJ...769...34V}. The standard rate equation
approach to the surface chemistry was utilized. Overall, the disk chemical network consists of 654 species
made of 12 elements, including dust grains, and 7299 reactions.

The age of the DM~Tau system is $\sim$ 3--7~Myr \citep[][]{Simon_ea00}, we used $5$~Myr in the chemical simulations.
The ``low metals'' elemental abundances of \citet{1982ApJS...48..321G,Lea98,2013ChRv1138710A} were utilized
(see Table~\ref{tabinit_abunds}). Parameters of the ``standard'' (or reference) DM Tau disk model
are summarized in Table~\ref{tabstdmodel}.

\begin{table}
\centering
\caption{Initial abundances used for the DM~Tau chemical modeling.}
\begin{tabular}{llllllllll}
\hline\hline
Species  & Relative abundances\\
\hline
H$_2$&   $0.499$     \\
H    &   $2.00 (-3)$  \\
He   &   $9.75 (-2)$  \\
C    &   $7.86 (-5)$  \\
N    &   $2.47 (-5)$  \\
O    &   $1.80 (-4)$  \\
S    &   $9.14 (-8)$  \\
Si   &   $9.74 (-9)$  \\
Na   &   $2.25 (-9)$  \\
Mg   &   $1.09 (-8)$  \\
Fe   &   $2.74 (-9)$  \\
P    &   $2.16 (-10)$ \\
Cl   &   $1.00 (-9)$  \\
\hline
\end{tabular}
\label{tabinit_abunds}
\end{table}

\begin{table}
\begin{center}
\caption{Parameters of the standard DM Tau disk model.\label{tabstdmodel}
}
\begin{tabular}{l|l|l}
\hline
\hline
Parameter & Symbol & Value \\
\hline
\multicolumn{3}{c}{Star}\\
\hline
Distance & $d_*$ & 140~pc\\
Temperature & $T_\mathrm{eff}$ & $3\,720$K \\
Radius & $R_*$ & $1.2R_\odot$ \\
Mass & $M_*$ & $0.65M_\odot$ \\
UV flux at 100~au & $\chi_*$ & $410\chi_{0}$ \\
X-ray luminosity$^{a}$ & $L_\mathrm{X}$ & $10^{31}$~erg\,s$^{-1}$\\
\hline
\multicolumn{3}{c}{Disk}\\
\hline
Accretion rate & $\dot{M}$ & $5\cdot10^{-9}M_\odot$\,yr$^{-1}$ \\
Viscosity  & $\alpha$ & $10^{-2}$ \\
Mass & $M_\mathrm{d}$ & $6.65\cdot10^{-2}M_\odot$ \\
Surface density at 1~au & $\Sigma_0$ & $4.182\cdot10^{24}$~cm$^{-2}$ \\
Inner radius & $r^\mathrm{d}_\mathrm{in}$ & $0.03$~au\\
Outer radius & $r^\mathrm{d}_\mathrm{out}$ & $800$~au\\
Gas-to-dust mass ratio & $m_\mathrm{gd}$ & 100\\
X-ray ionization exponent & $n$ & $2.81$\\
X-ray cross section at 1~keV & $\sigma_{-22}$ & $0.85 \times10^{-22}$~cm$^2$ \\
CRP ionization rate & $\zeta_\mathrm{CRP}$ & $1.3\cdot10^{-17}$~s$^{-1}$\\
SLR ionization rate & $\zeta_\mathrm{SLR}$ & $6.5\cdot10^{-19}$~s$^{-1}$\\
\hline
\multicolumn{3}{c}{Chemistry}\\
\hline
Average grain radius & $a_\mathrm{d}$ & $0.1\mu$m\\
Average grain density & $\rho_\mathrm{d}$ & $3$~g\,cm$^{-3}$\\
Sticking coefficient$^{b}$ & $S$ & $1$ \\
Amount of surface sites & $N_{s}$ & $1.88\times10^6$ \\
UV photodesorption yield & $Y_\mathrm{UV}$ & $10^{-5}$ \\
Chemical desorption yield & $Y_\mathrm{CD}$ & $10^{-2}$\\\hline
\end{tabular}
\tablefoot{$^{a}$ The X-ray emitting source assumed to be located at 12 stellar radii above the star.
$^{b}$ We assume that H$_2$ does not stick to dust grains.}
\end{center}
\end{table}

\section{Results}
\label{sec:results}

\begin{table*}
\centering
\caption{A grid of DM~Tau models is computed by varying key disk physical and chemical parameters.\label{tabruns}}
\begin{tabular}{l|lll}
\hline\hline
Parameter  & \multicolumn{3}{c}{Values}\\\cline{2-4}
           & Low & Medium & High \\
\hline
Turbulent mixing efficiency$^{*}$ & $\boldsymbol{Sc=\infty}$ & $Sc = 100$ & $Sc = 1$ \\
Average grain size & $\boldsymbol{0.1\mu\mathrm{m}}$ & $1.0\mu$m & $10.0\mu$m \\
$E_{\rm diff}/E_{\rm bind}$ & 0.3 & 0.5 & $\boldsymbol{0.77}$ \\
Initial S-abundance & $9\times10^{-9}$ & $\boldsymbol{9\times10^{-8}}$ & $9\times10^{-7}$\\
X-ray luminosity & $10^{29}$~erg\,s$^{-1}$ & $10^{30}$~erg\,s$^{-1}$ & $\boldsymbol{10^{31}~\mathrm{erg\,s}^{-1}}$ \\
Interstellar UV field     & $\boldsymbol{1\chi_{0}}$    & $10\chi_{0}$   & $100\chi_{0}$\\
CRP ionization & $1.3\times10^{-18}$~s$^{-1}$ & $\boldsymbol{1.3\times10^{-17}~\mathrm{s}^{-1}}$ & $1.3\times10^{-16}$~s$^{-1}$\\
C/O ratio & $\boldsymbol{0.46}$ & 1.0 & 1.2 \\
\hline
\end{tabular}
\tablefoot{$^{*}$ The parameters of the ``standard'' DM~Tau disk model are marked with boldface.}
\end{table*}

Using the DM~Tau disk physical and chemical model presented above, we performed about 20 detailed
chemical calculations by varying various physical and chemical parameters as shown in Table~\ref{tabruns}.
The results are compared to the reference model summarized in Table~\ref{tabstdmodel}.
The computed abundances of sulfur species at 5~Myr were vertically integrated to obtain radial distributions of the
column densities. After that, medians of these column densities were computed at radii between 100 and 800~au and compared
with the best-fit disk-averaged quantities derived with RADEX and the best-fit values at the radius of 300~au derived with DiskFit,
see Figs.~\ref{fig:sulfur_chem_transport}-\ref{fig:sulfur_chem_C2O} and related discussion.

{\it The key result of our study is that the observed CS column density and the sensitive upper limits for other S-bearing
species can only be matched when the gas has a non-solar metallicity with a C/O ratio of $\approx 1$, see Fig.~\ref{fig:sulfur_chem_C2O}}.
Another favorable model has a solar C/O ratio of 0.46 but requires a rather low X-ray luminosity of $\lesssim 10^{29}$~erg\,s$^{-1}$,
and a low initial sulfur abundance of $\lesssim 10^{-8}$ (Fig.~\ref{fig:sulfur_chem_LX}).
In all the other cases the observed and modeled CS/SO ratios disagree.

{\it Another key result of our modeling is that a combination of sulfur-bearing molecules, if detected in disks,
could provide very useful and unique diagnostics of a variety of disk physical and chemical processes.}
The summary of which sulfur-bearing species and what ratios are sensitive to what specific disk physical and chemical parameters
is presented in Table~\ref{tabouter_disk_tracers}. The CS/SO and CS/SO$_{2}$ ratios are mainly sensitive to the local
gas-phase C/O elemental ratio. The presence of abundant SO and SO$_{2}$ in the gas indicates either their more efficient surface
synthesis via fast diffusivity of S, O and other heavy radicals or more efficient release of SO, SO$_{2}$ ices
into the gas phase (shocks, high-energy irradiation, turbulent transport). In ideal circumstances when H$_{2}$S, OCS,
and H$_{2}$CS could also be observed in a disk, one could even determine which process is dominating.

\begin{table*}
\centering
\caption{\label{tabouter_disk_tracers} Sulfur-bearing tracers of outer disk parameters}
\begin{tabular}{llll}
\hline
\hline
Physical &  \multicolumn{3}{c}{Outer disk ($>100$~AU)}\\
processes & Correlate & Do not correlate & Anti-correlate \\
\hline
C/O ratio  & CS, CCS  & OCS, H$_2$CS, H$_2$S & SO, SO$_2$ \\
\hline
Surface diffusivity  & SO$_2$, OCS & SO, CS, H$_2$S & CCS, H$_2$CS\\
\hline
Turbulent mixing   & CS, SO, SO$_2$, OCS, CS, CCS &  H$_2$CS, H$_2$S & ... \\
\hline
X-ray luminosity & CCS$^*$ & SO, SO$_2$, OCS, H$_2$CS, H$_2$S & CS \\
\hline
UV intensity & SO, SO$_2$, H$_2$S &  OCS, CS, CCS, H$_2$CS  & ... \\
\hline
Grain growth & CS, SO, SO$_2$, CCS & OCS, H$_2$CS, H$_2$S & ... \\
\hline
\end{tabular}
\tablefoot{$^{*}$ The CCS column density first decreases when the X-ray luminosity
increases from $10^{29}$ to $10^{30}$~erg\,s$^{-1}$ but then it increases when the
X-ray luminosity further increases to $10^{31}$~erg\,s$^{-1}$.}
\end{table*}

\subsection{Impact of turbulent mixing}
\label{sec:transport}

\begin{figure}
\centering
\includegraphics[width=0.98\hsize]{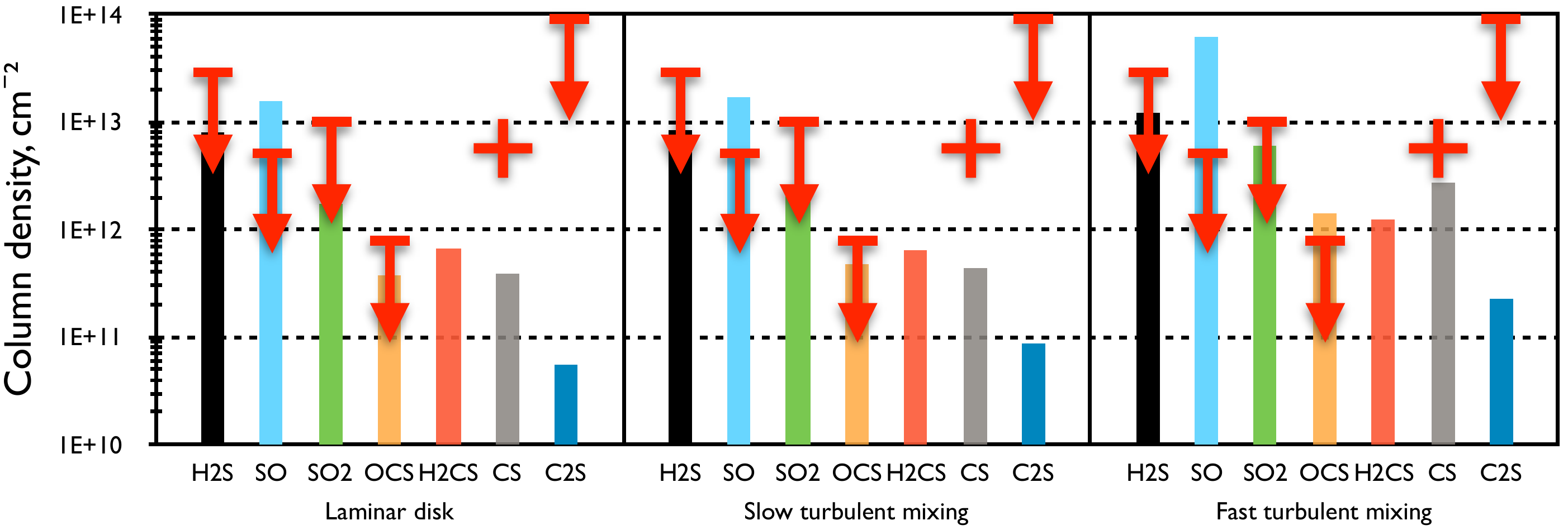}
\caption{\label{fig:sulfur_chem_transport}The observed and computed median column densities of the key sulfur-bearing species at $r=100-800$~au
in the DM~Tau disk at 5~Myr are compared. Cross indicates the CS column density and the arrows indicate the upper limits of the observed column densities
for other S-bearing molecules. 
(Left to right) Three scenarios of turbulent mixing are considered: (1) laminar disk ($Sc=\infty$), (2) slow mixing ($Sc=100$), and (3) fast mixing ($Sc=1$) \citep{SW2011}.}
\end{figure}

The turbulent transport in disks enriches chemistry by bringing chemical ingredients from cold and dark
outer regions to those where reactions with barriers, ice evaporation and mild high-energy processing of gas and ices
become active \citep[see e.g.][]{IHMM04,Heinzeller_ea11,SW2011,2014ApJ...790...97F}. This is particularly true for chemical species
that are not volatile and/or produced at least partly by surface reactions, including sulfurette molecules.
As a result of transport, all sulfur species show a uniform increase in column densities by $\lesssim 3-10$,
especially, when the mixing is fast (see Fig.~\ref{fig:sulfur_chem_transport}). Consequently, such an uniform increase does not strongly
affect the computed CS/SO ratio, which remains similar in the laminar (CS/SO $\approx 0.026$) and the fast turbulent mixing (CS/SO $\approx 0.044$) cases.
Both these values are too low compared with our observations showing CS/SO $\gtrsim 1$.

\subsection{Impact of initial sulfur abundances}
\label{sec:init_abunds}

\begin{figure}
\centering
\includegraphics[width=0.98\hsize]{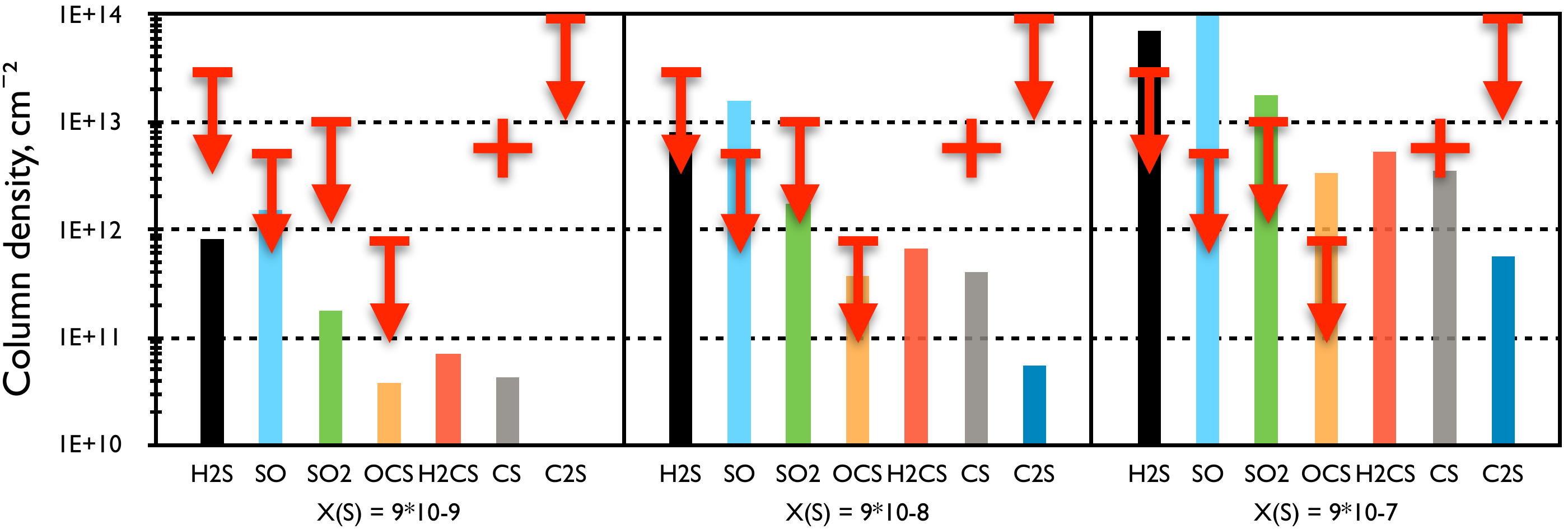}
\caption{\label{fig:sulfur_chem_initial_abundance} As Fig.~\ref{fig:sulfur_chem_transport} but for the sulfur initial elemental abundances.
Cross indicates the CS column density and the arrows indicate the upper limits of the observed column densities
for other S-bearing molecules. 
(Left to right) Three sets of initial elemental sulfur abundances are considered: (1) $X({\rm S})=9\times10^{-9}$,
(2) $X({\rm S})=9\times10^{-8}$, and (3) $X({\rm S})=9\times10^{-7}$.}
\end{figure}

Another crucial factor impacting the outcome of sulfur chemistry is the total amount of elemental sulfur available for chemistry.
The higher the initial sulfur elemental abundance that is not locked inside dust grains, the higher the
computed abundances and column densities for the sulfur molecules and ices. The response of disk sulfur chemistry to
the linear increase of the initial sulfur elemental abundance in our model is almost linear, see Fig.~\ref{fig:sulfur_chem_initial_abundance}.
This again means that the resulting CS/SO ratio remains too low, $\sim 0.02$, as compared to our interferometric observations showing
CS/SO $\gtrsim 1$.

\subsection{Impact of surface diffusivities}
\label{sec:surf_diff}

\begin{figure}
\centering
\includegraphics[width=0.98\hsize]{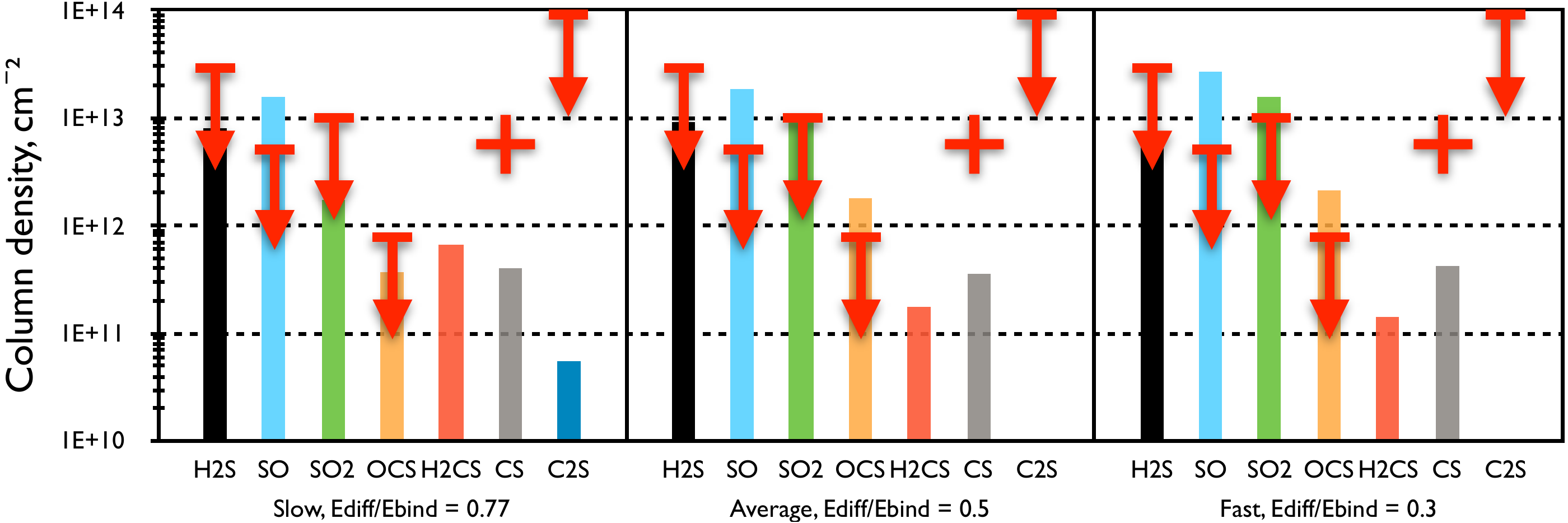}

\caption{\label{fig:sulfur_chem_diffusivity} As Fig.~\ref{fig:sulfur_chem_transport} but for the surface diffusivities.
Cross indicates the CS column density and the arrows indicate the upper limits of the observed column densities
for other S-bearing molecules. 
(Left to right) Three scenarios for the surface diffusivity are considered: (1) slow diffusion, $E_{\rm diff}/E_{bind} = 0.77$,
(2) average diffusion, $E_{\rm diff}/E_{bind} = 0.5$, and (3) fast diffusion, $E_{\rm diff}/E_{bind} = 0.3$.}
\end{figure}

Unlike the previous two parameters, the pace of surface chemistry regulated by the surface diffusivities affects the disk sulfur chemistry
differently (Fig.~\ref{fig:sulfur_chem_diffusivity}). SO$_{2}$, OCS, and to a lesser degree SO
can be efficiently produced via surface processes involving S, O, CO or SO only when these heavy ices are mobile and hence
benefit from the fast, $E_{\rm diff}/E_{bind} = 0.3$ diffusion. The H$_{2}$S that is also produced on
grains does not benefit that much from faster diffusivities because it is formed via direct hydrogenation of S, and H atoms can
rapidly scan the dust surfaces even at $T=10-20$~K and when the assumed diffusivity is slow, $E_{\rm diff}/E_{\rm bind} = 0.77$.
Other S-bearing species show two distinct trends. While CS and partly SO, synthesized mainly in the gas phase, are not strongly affected by the
pace of surface chemistry, the column densities of C$_{2}$S and H$_{2}$CS decline when the surface diffusivities increase.
For these two species the decline is caused by the competition between the two main C-reservoirs, CO and CH$_{4}$.
The synthesis of CH$_{4}$ greatly benefits from faster diffusion, which lowers gas-phase abundances of other simple hydrocarbons like CH, CH$_{3}$, etc.
out of which C$_{2}$S and H$_{2}$CS are produced.

As above, the modeled CS/SO ratios when the surface diffusivities are varied remain too low compared to the observations.

\subsection{Impact of grain growth}
\label{sec:grain_growth}

\begin{figure}
\centering
\includegraphics[width=0.98\hsize]{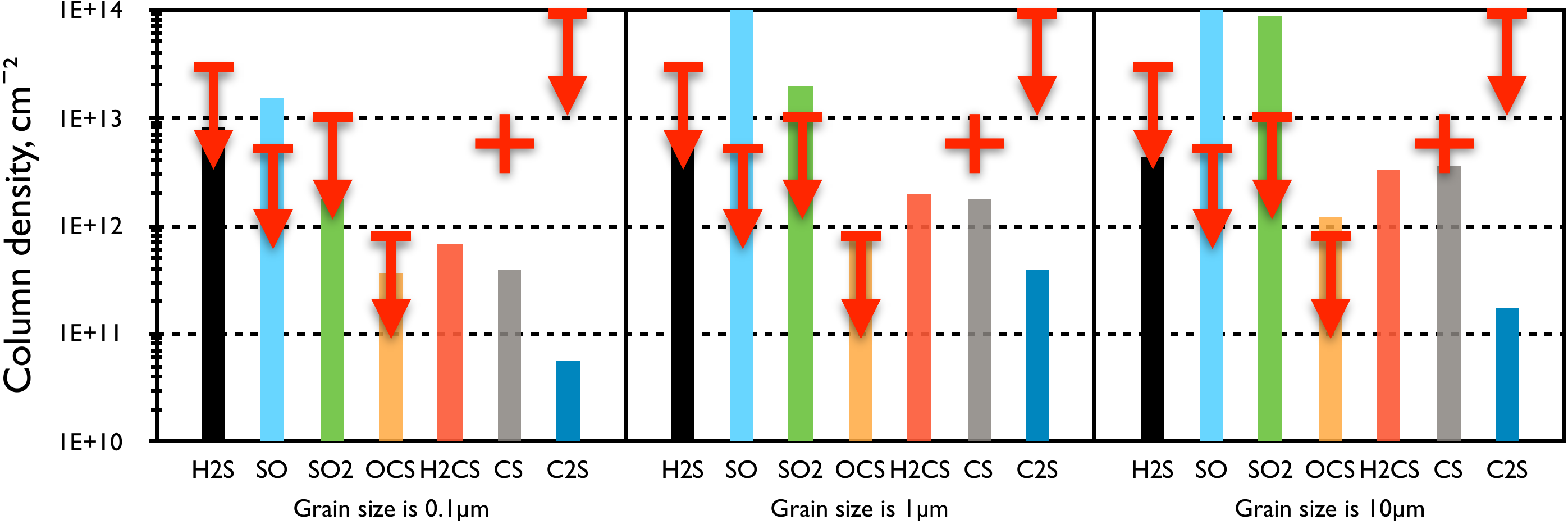}

\caption{\label{fig:sulfur_chem_grain_evolution}As Fig.~\ref{fig:sulfur_chem_transport} but for the grain sizes.
Cross indicates the CS column density and the arrows indicate the upper limits of the observed column densities
for other S-bearing molecules. 
(Left to right) Three uniform dust grain sizes are considered: (1) $0.1 {\mu}m$, (2) $1 {\mu}m$, and (3) $10 {\mu}m$.}
\end{figure}

The main effect of grain growth on disk chemistry, when the total dust/gas ratio does not severely change, is the
reduction of the grain surface available for accretion (per unit gas volume). The disk sulfur chemistry reacts
to the uniform grain growth from the ISM-like $0.1 \mu$m-sized grains to the $10 \mu$m-sized grains by increasing
abundances and column densities of nearly all S-species due to their less severe freeze-out and more efficient gas-phase chemistry
with S, particularly, for
SO and SO$_{2}$, see Fig.~\ref{fig:sulfur_chem_grain_evolution}. The
only species that shows an opposite trend is H$_{2}$S, which column densities decrease by a small factor of 2
(from $8\times10^{12}$~cm$^{-2}$ to $4\times10^{12}$~cm$^{-2}$) with grain growth.
This species is produced almost entirely by the surface hydrogenation of
sulfur and its production efficiency suffers from the reduced grain surface available for this process.
Grain growth, as the other disk parameters discussed above, does not bring the modeled CS/SO ratios of $\approx 0.026$,
$\approx 0.018$, and $\approx 0.009$ (for the $0.1$, $1$, and $10 \mu$m-sized grains, respectively) to the observed value
of $\gtrsim 1$.

\subsection{Impact of stellar X-ray luminosity}
\label{sec:LX}

\begin{figure}
\centering
\includegraphics[width=0.98\hsize]{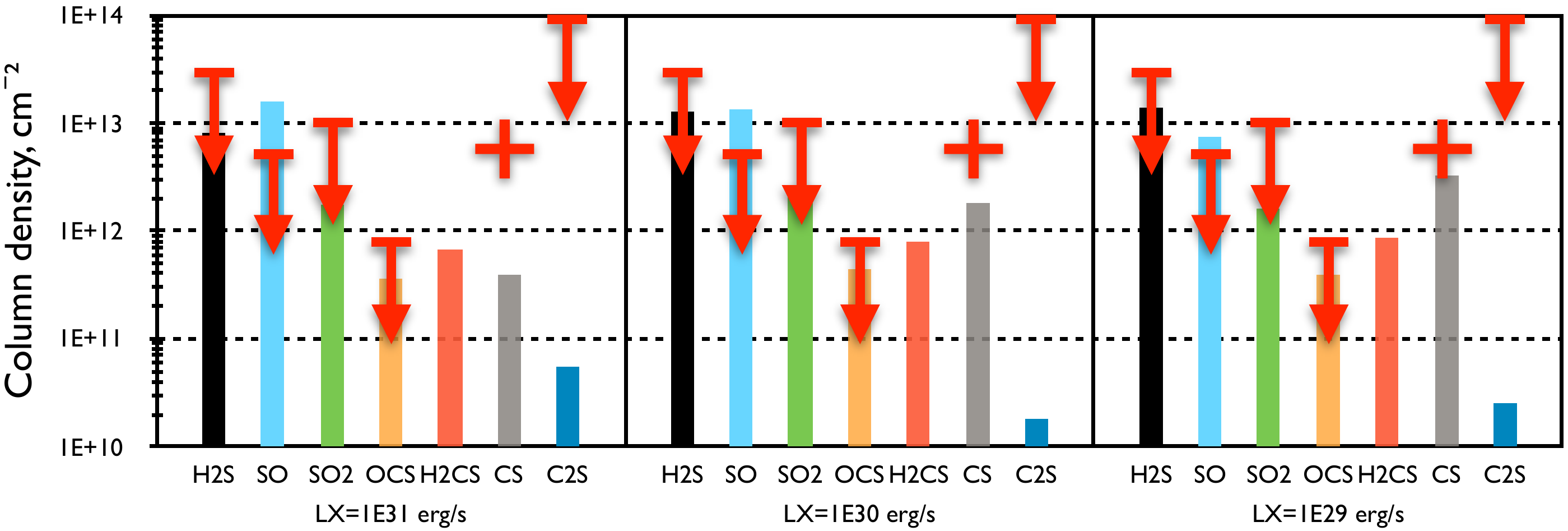}

\caption{\label{fig:sulfur_chem_LX}
As Fig.~\ref{fig:sulfur_chem_transport} but for the stellar X-ray luminosities.
Cross indicates the CS column density and the arrows indicate the upper limits of the observed column densities
for other S-bearing molecules. 
(Left to right) Three cases of the stellar X-ray luminosities are considered: (1) $L_{X} = 10^{31}$~erg\,s$^{-1}$,
(2) $L_{X} = 10^{30}$~erg\,s$^{-1}$, and (3) $L_{X} = 10^{29}$~erg\,s$^{-1}$.}
\end{figure}

The change of the stellar X-ray luminosity affects chemistry differently for CS and SO and other sulfur species,
see Fig.~\ref{fig:sulfur_chem_LX}.
When the adopted $L_{X}$ decreases from $10^{31}$~erg\,s$^{-1}$ to $10^{29}$~erg\,s$^{-1}$, the corresponding
CS column densities increase from $\approx 4\times 10^{11}$ to $3\times 10^{12}$~cm$^{-2}$. Contrary,
the SO column densities decrease by a factor of 2 from $\approx 1.5\times 10^{13}$ to $\sim 8\times 10^{12}$~cm$^{-2}$.
The SO abundances decrease with decreasing X-ray ionization rate because 1) less SO molecules can be kicked out from
dust grain surfaces and 2) because less amount of oxygen can be liberated from O-bearing ices and from gaseous CO
upon destruction by the X-ray-produced He$^{+}$ ions. The CS, in contrast, benefits from extra carbon liberated from
CO by He$^{+}$.

Consequently, the modeled CS/SO ratio increases from $\approx 0.026$ to 0.43, getting closer to the observed value
of $\gtrsim 1$. The observed CS column density and the SO upper limits derived with DiskFit are matched with the model
with the ``standard'' S elemental abundance of $9\times 10^{-8}$ (Table~\ref{tabinit_abunds}).
In contrast, in order to better match the absolute values of the observed CS column density and the SO upper limits
derived with RADEX, the model with the low stellar X-ray luminosity of $10^{29}$~erg\,s$^{-1}$ needs an additional
downscaling of the initial sulfur elemental abundance by a factor of 5, to the value of $\approx 2\times 10^{-8}$.

This is the first model that fits the data and the observed CS/SO ratio, although the
best-fit stellar X-ray luminosity
of $10^{29}$~erg\,s$^{-1}$ seems to be to low compared to the observed X-ray properties of DM~Tau.
Based on the X-ray measurements with {\it Chandra} and {\it XMM} in the range of $0.3-10$~keV, the quiescent stellar X-ray luminosity of
DM~Tau is about $10^{30}$~erg\,s$^{-1}$ (M.~Guedel, private communication), and can temporarily rise during the flares.
This value is close to the median value representative of the X-ray luminosities measured in T~Tauri stars, $10^{29}-10^{31}$~erg\,s$^{-1}$
\citep[e.g.,][]{FeigelsonMontmerle99, Telleschi_ea07}. In the previous chemical studies of the DM~Tau disk comparable values of $2-100 \times 10^{29}$~erg\,s$^{-1}$ were used \citep[][]{SW2011,Cleeves:2014bn,Teague:2015jk,Bergin:2016ge,Rab_ea17a}.

\subsection{Impact of cosmic ray ionization}
\label{sec:CRP}

\begin{figure}
\centering
\includegraphics[width=0.98\hsize]{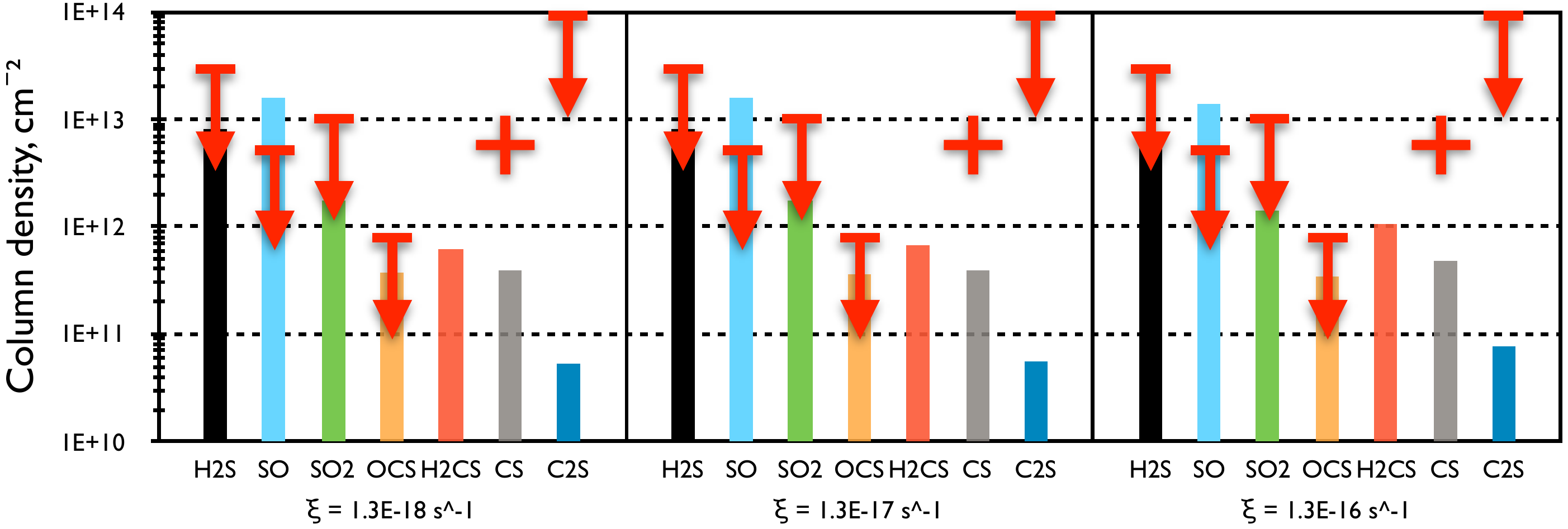}

\caption{\label{fig:sulfur_chem_CRP}
As Fig.~\ref{fig:sulfur_chem_transport} but for the cosmic ray ionization rates.
Cross indicates the CS column density and the arrows indicate the upper limits of the observed column densities
for other S-bearing molecules. 
(Left to right) Three CRP ionization rates were considered: (1) $\zeta_{\rm CRP}= 1.3\times10^{-18}$~s$^{-1}$,
(2)  $\zeta_{\rm CRP}= 1.3\times10^{-17}$~s$^{-1}$, and (3) $\zeta_{\rm CRP}= 1.3\times10^{-16}$~s$^{-1}$.}
\end{figure}

The impact of the cosmic ray ionization rate on the disk sulfur chemistry is relatively minimal, see Fig.~\ref{fig:sulfur_chem_CRP}.
The most affected species, H$_{2}$CS, shows an increase in the column densities from $6\times 10^{11}$ to
$\approx 10^{12}$~cm$^{-2}$ for the low and high CRP ionization rates of $1.3\times10^{-18}$~s$^{-1}$ and $1.3\times10^{-16}$~s$^{-1}$, respectively.
As we mentioned in the Section~\ref{sec:disk_chem}, in our disk model the X-ray-driven ionization dominates over the CRP ionization
in the disk molecular layer, where gaseous sulfur-bearing molecules have peak abundances.
Thus, by varying CRP ionization rates the modeled CS/SO ratios remain too low, $\lesssim 0.035$, compared to the observed value
of $\gtrsim 1$.

\subsection{Impact of interstellar UV radiation}
\label{sec:ISUV}

\begin{figure}
\centering
\includegraphics[width=0.98\hsize]{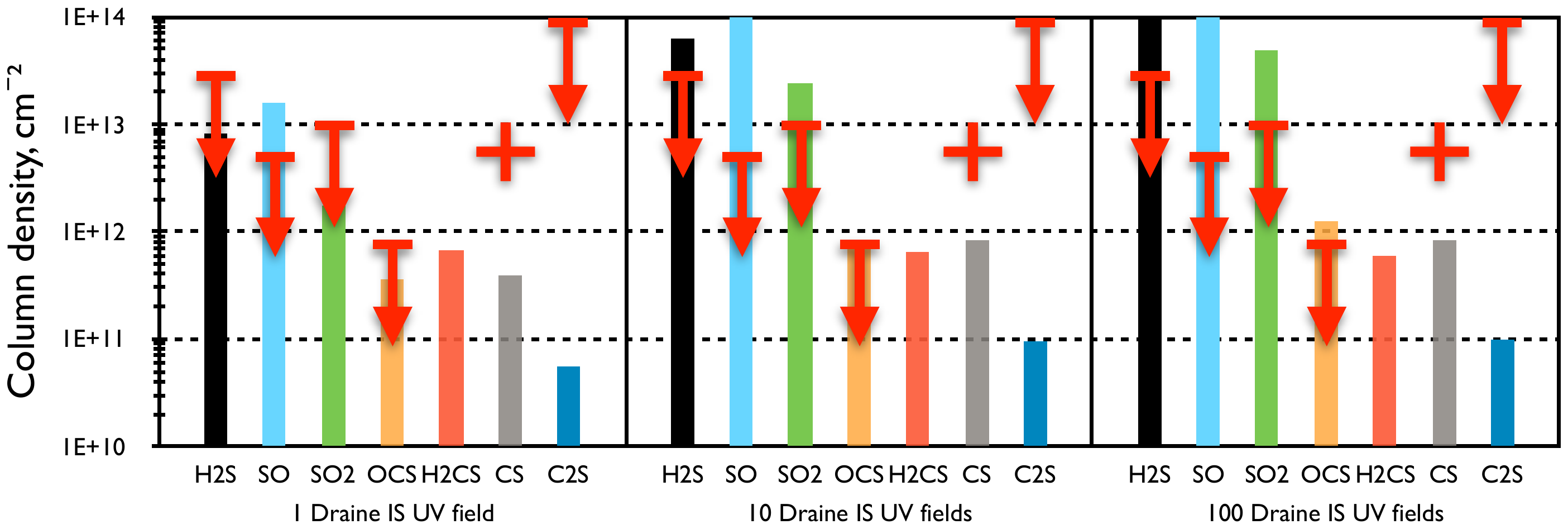}
\caption{\label{fig:sulfur_chem_LISUV}
As Fig.~\ref{fig:sulfur_chem_transport} but for the interstellar UV radiation fields.
Cross indicates the CS column density and the arrows indicate the upper limits of the observed column densities
for other S-bearing molecules. 
(Left to right) Three cases of the interstellar UV radiation field are considered: (1) $1\chi_{0}$,
(2) $10\chi_{0}$, and (3) $100\chi_{0}$, where $\chi_{0}$ is the Draine~(1978) IS UV field.}
\end{figure}

The interstellar UV radiation regulates photodissociation and photoionization of gas as well as photodesorption and photoprocessing
of ices. The key process for the chemistry in the disk molecular layer is photodesorption of ices. When more UV radiation
penetrates into this layer, it liberates refractory ices more efficiently, increasing their
gas-phase concentrations. In addition, more energetic UV-irradiation of ices create reactive radicals including O and S,
boosting the surface synthesis of chemically-related molecular ices.

This is what can be seen in Fig.~\ref{fig:sulfur_chem_LISUV}. The heavy molecules
that are mainly synthesized on grains, like SO$_{2}$, OCS and H$_{2}$S show higher column densities
by factors of $\sim 4-25$ for the case when the IS UV field is $\chi_{\rm UV}= 100\chi_{0}$ compared to the standard $\chi_{\rm UV}= 1\chi_{0}$ case.
In contrast, CS, H$_{2}$CS, and CCS are produced mainly in the gas phase and hence their
abundances are not that sensitive to changes of the UV radiation intensity.
Consequently, by increasing the amount of the UV radiation penetrating into the DM~Tau molecular layer,
the resulting CS/SO ratios can only become lower, from $\approx 0.026$ for the $\chi_{\rm UV}= 1\chi_{0}$ model to
$\approx 0.004$ for the $\chi_{\rm UV}= 100\chi_{0}$ model.

\subsection{Impact of C/O ratios}
\label{sec:C2O}

\begin{figure}
\centering
\includegraphics[width=0.98\hsize]{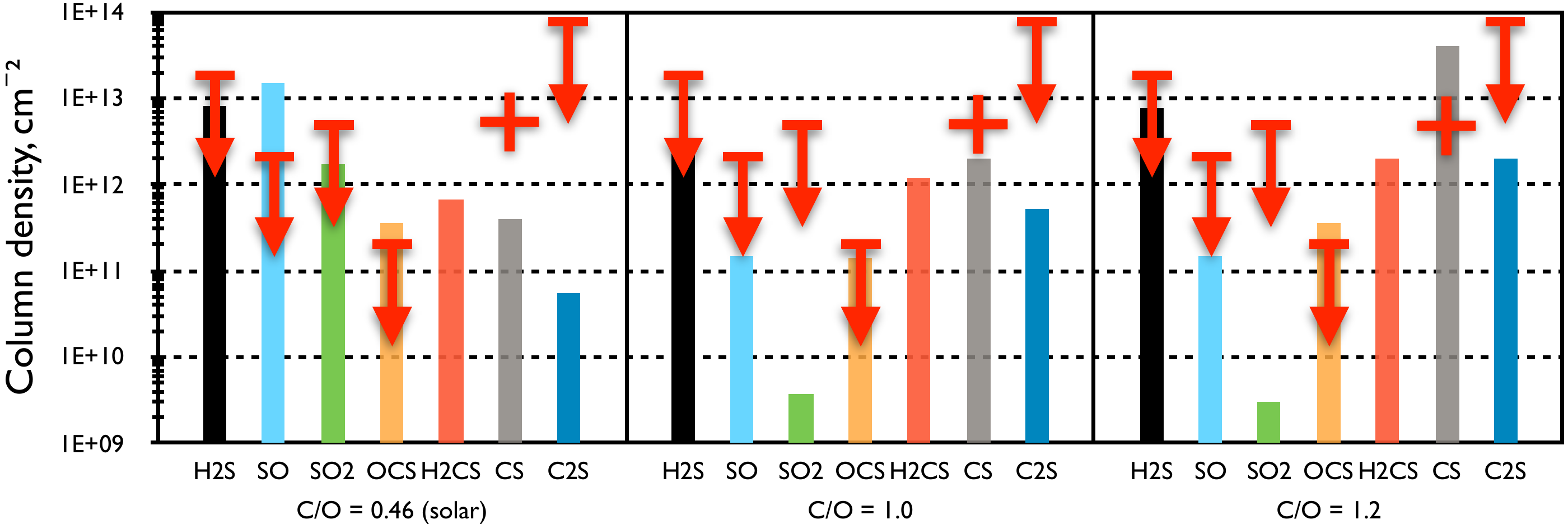}
\caption{\label{fig:sulfur_chem_C2O}
As Fig.~\ref{fig:sulfur_chem_transport} but for the elemental C/O ratios.
Cross indicates the CS column density and the arrows indicate the upper limits of the observed column densities
for other S-bearing molecules. 
(Left to right) Three elemental C/O ratios are modeled: (1) the solar C/O ratio of 0.46,
(2) the C/O ratio of 1.0, and (3) the C/O ratio of 1.2.}
\end{figure}

One of the most important parameters shaping the entire disk chemistry is the total elemental C/O ratio as well
as the degree of carbon and oxygen depletion. As soon as the C/O ratio switches from $<1$ to $>1$, the abundances of
many oxygen-bearing species drop substantially, as the majority of elemental O becomes locked in CO. In contrast,
the concentrations of carbon-bearing species increase due to the availability of elemental C that is not bound in CO.
This makes the CS/SO and CS/SO$_{2}$ ratios extremely sensitive to the local disk C/O ratio, see Fig.~\ref{fig:sulfur_chem_C2O}.

The column densities of SO and SO$_{2}$ in our DM~Tau model drop by factors of $\sim 100$ and 500, respectively,
when the C/O ratio changes from 0.46 to 1.2. Contrary, the column densities of CS, CCS, and H$_{2}$CS are increased by
factors of $\sim 100$, 35, 2, respectively. The column densities of H$_{2}$S do not change as it contains
neither C nor O. The column densities of OCS also do not strongly change because this molecule has an efficient
surface formation route via CO + S, and hence follows the CO evolution.

As a result, the modeled CS/SO ratio increases from the ``standard'' low value of $\approx 0.026$ (C/O $=0.46$) to
the values of 13 (C/O $=1.0$) and 282 (C/O $=1.2$). The trends continues with increasing the C/O ratio,
for C/O $=2$ the modeled CS/SO ratio is $498$. As can be clearly seen in Figure~\ref{fig:sulfur_chem_C2O},
the modeled and observed data agree reasonably well (taking modeling uncertainties of a factor of $\sim 3-10$ into account).

\subsection{Analysis of sulfur chemistry}
\label{sec:analysis}
\longtab{
\begin{longtable}{lrrr}
\caption{\label{tabkey_reac_S} Key chemical processes for S-bearing species}\\
\hline
\hline
Reaction & $\alpha$ & $\beta$ & $\gamma$ \\
         & (cm$^3$)\,s$^{-1}$ & -- & (K) \\
\hline
\endfirsthead
\caption{continued.}\\
\hline
\hline
Reaction & $\alpha$ & $\beta$ & $\gamma$ \\
         & (cm$^3$)\,s$^{-1}$ & -- & (K) \\
\hline
\endhead
\hline
\endfoot
\multicolumn{4}{c}{Ion-molecule reactions}\\
\hline
CH$_{3}$ + S$^{+}$  $\rightarrow$  HCS$^{+}$ + H$_{2}$   & $5.00\,(-10)$   & $0$   & $0$   \\
CH$_3$$^+$   + S  $\rightarrow$  HCS$^+$   + H$_2$   & $0.14\,(-8)$   & $0$   & $0$   \\
S$^+$   + C$_2$H  $\rightarrow$  C$_2$S$^+$   + H   & $0.20\,(-8)$   & $-0.50$   & $0$   \\
S$^+$   + C$_3$H  $\rightarrow$  C$_3$S$^+$   + H   & $0.69\,(-8)$   & $-0.50$   & $0$   \\
S$^+$   + CH$_4$  $\rightarrow$  H$_3$CS$^+$   + H   & $0.14\,(-9)$   & $0$   & $0$   \\
CS$^+$   + H$_2$  $\rightarrow$  HCS$^+$   + H   & $0.48\,(-9)$   & $0$   & $0$   \\
H$^+$   + C$_2$S  $\rightarrow$  C$_2$S$^+$   + H   & $0.11\,(-7)$   & $-0.50$   & $0$   \\
H$^+$   + C$_3$S  $\rightarrow$  C$_3$S$^+$   + H   & $0.15\,(-7)$   & $-0.50$   & $0$   \\
CS   + H$^+$  $\rightarrow$  CS$^+$   + H   & $0.18\,(-7)$   & $-0.50$   & $0$   \\
H$_2$S   + H$^+$  $\rightarrow$  H$_2$S$^+$   + H   & $0.38\,(-8)$   & $-0.50$   & $0$   \\
OCS   + H$^+$  $\rightarrow$  HS$^+$   + CO   & $0.65\,(-8)$   & $-0.50$   & $0$   \\
SO   + H$^+$  $\rightarrow$  SO$^+$   + H   & $0.14\,(-7)$   & $-0.50$   & $0$   \\
CS   + H$_3$$^+$  $\rightarrow$  HCS$^+$   + H$_2$   & $0.11\,(-7)$   & $-0.50$   & $0$   \\
H$_3$$^+$   + H$_2$CS  $\rightarrow$  H$_3$CS$^+$   + H$_2$   & $0.38\,(-8)$   & $-0.50$   & $0$   \\
H$_2$S   + H$_3$$^+$  $\rightarrow$  H$_3$S$^+$   + H$_2$   & $0.23\,(-8)$   & $-0.50$   & $0$   \\
OCS   + H$_3$$^+$  $\rightarrow$  HOCS$^+$   + H$_2$   & $0.38\,(-8)$   & $-0.50$   & $0$   \\
SO   + H$_3$$^+$  $\rightarrow$  HSO$^+$   + H$_2$   & $0.84\,(-8)$   & $-0.50$   & $0$   \\
SO$_2$   + H$_3$$^+$  $\rightarrow$  HSO$_2$$^+$   + H$_2$   & $0.37\,(-8)$   & $-0.50$   & $0$   \\
CS   + H$_3$O$^+$  $\rightarrow$  HCS$^+$   + H$_2$O   & $0.50\,(-8)$   & $-0.50$   & $0$   \\
H$_2$S   + H$_3$O$^+$  $\rightarrow$  H$_3$S$^+$   + H$_2$O   & $0.11\,(-8)$   & $-0.50$   & $0$   \\
CS   + HCO$^+$  $\rightarrow$  HCS$^+$   + CO   & $0.43\,(-8)$   & $-0.50$   & $0$   \\
HCO$^+$   + H$_2$CS  $\rightarrow$  H$_3$CS$^+$   + CO   & $0.15\,(-8)$   & $-0.50$   & $0$   \\
H$_2$S   + HCO$^+$  $\rightarrow$  H$_3$S$^+$   + CO   & $0.95\,(-9)$   & $-0.50$   & $0$   \\
OCS   + HCO$^+$  $\rightarrow$  HOCS$^+$   + CO   & $0.15\,(-8)$   & $-0.50$   & $0$   \\
SO   + HCO$^+$  $\rightarrow$  HSO$^+$   + CO   & $0.33\,(-8)$   & $-0.50$   & $0$   \\
HCS$^+$   + O  $\rightarrow$  OCS$^+$   + H   & $0.50\,(-9)$   & $0$   & $0$   \\
HCS$^+$   + O  $\rightarrow$  HCO$^+$   + S   & $0.50\,(-9)$   & $0$   & $0$   \\
H$_2$S   + S$^+$  $\rightarrow$  S$_2$$^+$   + H$_2$   & $0.64\,(-9)$   & $-0.50$   & $0$   \\
OCS   + S$^+$  $\rightarrow$  S$_2$$^+$   + CO   & $0.91\,(-9)$   & $-0.50$   & $0$   \\
C$^+$   + C$_2$S  $\rightarrow$  C$_2$S$^+$   + C   & $0.17\,(-8)$   & $-0.50$   & $0$   \\
C$^+$   + C$_3$S  $\rightarrow$  C$_3$S$^+$   + C   & $0.24\,(-8)$   & $-0.50$   & $0$   \\
C$^+$   + H$_2$CS  $\rightarrow$  CS   + CH$_2$$^+$   & $0.21\,(-8)$   & $-0.50$   & $0$   \\
H$_2$S   + C$^+$  $\rightarrow$  HCS$^+$   + H   & $0.95\,(-9)$   & $-0.50$   & $0$   \\
OCS   + C$^+$  $\rightarrow$  CS$^+$   + CO   & $0.16\,(-8)$   & $-0.50$   & $0$   \\
OCS   + C$^+$  $\rightarrow$  OCS$^+$   + C   & $0.40\,(-9)$   & $-0.50$   & $0$   \\
SO$_2$   + C$^+$  $\rightarrow$  SO$^+$   + CO   & $0.20\,(-8)$   & $-0.50$   & $0$   \\
\hline
\multicolumn{4}{c}{Dissociative recombination reactions}\\
\hline
C$_2$S$^+$   + e$^-$  $\rightarrow$  CS   + C   & $0.15\,(-6)$   & $-0.50$   & $0$   \\
C$_2$S$^+$   + e$^-$  $\rightarrow$  C$_2$   + S   & $0.15\,(-6)$   & $-0.50$   & $0$   \\
C$_3$S$^+$   + e$^-$  $\rightarrow$  C$_2$S   + C   & $1.00\,(-7)$   & $-0.50$   & $0$   \\
C$_3$S$^+$   + e$^-$  $\rightarrow$  C$_3$   + S   & $1.00\,(-7)$   & $-0.50$   & $0$   \\
C$_3$S$^+$   + e$^-$  $\rightarrow$  CS   + C$_2$   & $1.00\,(-7)$   & $-0.50$   & $0$   \\
H$_3$CS$^+$   + e$^-$  $\rightarrow$  H$_2$CS   + H   & $0.30\,(-6)$   & $-0.50$   & $0$   \\
H$_3$S$^+$   + e$^-$  $\rightarrow$  H$_2$S   + H   & $0.30\,(-6)$   & $-0.50$   & $0$   \\
HC$_2$S$^+$   + e$^-$  $\rightarrow$  CS   + CH   & $0.15\,(-6)$   & $-0.50$   & $0$   \\
HCS$^+$   + e$^-$  $\rightarrow$  CS   + H   & $0.18\,(-6)$   & $-0.57$   & $0$   \\
HCS$^+$   + e$^-$  $\rightarrow$  CH   + S   & $0.79\,(-6)$   & $-0.57$   & $0$   \\
HOCS$^+$   + e$^-$  $\rightarrow$  CS   + OH   & $0.20\,(-6)$   & $-0.50$   & $0$   \\
HOCS$^+$   + e$^-$  $\rightarrow$  OCS   + H   & $0.20\,(-6)$   & $-0.50$   & $0$   \\
HSO$^+$   + e$^-$  $\rightarrow$  SO   + H   & $0.20\,(-6)$   & $-0.50$   & $0$   \\
HSO$_2$$^+$   + e$^-$  $\rightarrow$  SO$_2$   + H   & $0.20\,(-6)$   & $-0.50$   & $0$   \\
HSO$_2$$^+$   + e$^-$  $\rightarrow$  SO   + OH   & $1.00\,(-7)$   & $-0.50$   & $0$   \\
OCS$^+$   + e$^-$  $\rightarrow$  CO   + S   & $0.29\,(-6)$   & $-0.62$   & $0$   \\
OCS$^+$   + e$^-$  $\rightarrow$  CS   + O   & $0.48\,(-7)$   & $-0.62$   & $0$   \\
OCS$^+$   + e$^-$  $\rightarrow$  SO   + C   & $0.10\,(-7)$   & $-0.62$   & $0$   \\
H$_2$CS$^+$   + e$^-$  $\rightarrow$  CS   + H   + H   & $0.30\,(-6)$   & $-0.50$   & $0$   \\
H$_3$CS$^+$   + e$^-$  $\rightarrow$  CS   + H   + H$_2$   & $0.30\,(-6)$   & $-0.50$   & $0$   \\
\hline
\multicolumn{4}{c}{Neutral-neutral reactions without barriers}\\
\hline
SO   + C  $\rightarrow$  CS   + O   & $0.35\,(-10)$   & $0$   & $0$   \\
SO   + C  $\rightarrow$  CO   + S   & $0.35\,(-10)$   & $0$   & $0$   \\
SO$_2$   + C  $\rightarrow$  CO   + SO   & $0.70\,(-10)$   & $0$   & $0$   \\
CS   + CH  $\rightarrow$  C$_2$S   + H   & $1.00\,(-10)$   & $0$   & $0$   \\
O   + HCS  $\rightarrow$  OCS   + H   & $0.50\,(-10)$   & $0$   & $0$   \\
O   + HS  $\rightarrow$  SO   + H   & $0.16\,(-9)$   & $0.50$   & $0$   \\
OH   + SO  $\rightarrow$  SO$_2$   + H   & $0.86\,(-10)$   & $0$   & $0$   \\
S   + C$_2$  $\rightarrow$  CS   + C   & $1.00\,(-10)$   & $0$   & $0$   \\
S   + CH  $\rightarrow$  CS   + H   & $0.50\,(-10)$   & $0$   & $0$   \\
S   + CH$_2$  $\rightarrow$  CS   + H$_2$   & $1.00\,(-10)$   & $0$   & $0$   \\
S   + O$_2$  $\rightarrow$  SO   + O   & $0.23\,(-11)$   & $0$   & $0$   \\
S   + OH  $\rightarrow$  SO   + H   & $0.66\,(-10)$   & $0$   & $0$   \\
%
\hline
\multicolumn{4}{c}{Surface photoprocessing (CRP-induced UV photons)}\\
\hline
H$_2$S (ice)   + h$\nu_{\rm CRP}$  $\rightarrow$  H$_2$ (ice)   + S (ice)   & $5.15\,(3)$   & $0$   & $0$   \\
H$_2$S (ice)   + h$\nu_{\rm CRP}$  $\rightarrow$  HS (ice)   + H (ice)   & $8.50\,(2)$   & $0$   & $0$   \\
OCS (ice)   + h$\nu_{\rm CRP}$  $\rightarrow$  CO (ice)   + S (ice)   & $5.35\,(3)$   & $0$   & $0$   \\
SO (ice)   + h$\nu_{\rm CRP}$  $\rightarrow$  S (ice)   + O (ice)   & $5.00\,(2)$   & $0$   & $0$   \\
SO$_2$ (ice)   + h$\nu_{\rm CRP}$  $\rightarrow$  SO (ice)   + O (ice)   & $1.88\,(3)$   & $0$   & $0$   \\
\hline
\multicolumn{4}{c}{Surface recombination reactions}\\
\hline
H (ice)   + S (ice)  $\rightarrow$  HS (ice)   & $1.00$   & $0$   & $0$   \\
H (ice)   + HS (ice)  $\rightarrow$  H$_2$S (ice)   & $1.00$   & $0$   & $0$   \\
O (ice)   + CS (ice)  $\rightarrow$  OCS (ice)   & $1.00$   & $0$   & $0$   \\
O (ice)   + S (ice)  $\rightarrow$  SO (ice)   & $1.00$   & $0$   & $0$   \\
SO (ice)   + O (ice)  $\rightarrow$  SO$_2$ (ice)   & $1.00$   & $0$   & $0$   \\
SO (ice)   + C (ice)  $\rightarrow$  CO (ice)   + S (ice)   & $1.00$   & $0$   & $0$   \\
S (ice)   + CO (ice)  $\rightarrow$  OCS (ice)   & $1.00$   & $0$   & $0$   \\
OCS (ice)   + H (ice)  $\rightarrow$  CO (ice)   + HS (ice)   & $1.00$   & $0$   & $0$   \\
SO$_2$ (ice)   + H (ice)  $\rightarrow$  O$_2$ (ice)   + HS (ice)   & $1.00$   & $0$   & $0$   \\
O (ice)   + HS (ice)  $\rightarrow$  SO (ice)   + H (ice)   & $1.00$   & $0$   & $0$   \\

\hline
\end{longtable}
}

We used our chemical analysis
software to identify the key gas-phase and surface formation and destruction pathways for CS, SO, SO$_{2}$, OCS, and H$_{2}$S at a
representative outer disk radius of 230~au of the ``standard'' DM~Tau model with the solar C/O elemental composition, as in \citet{SW2011}.
At this radius the temperature ranges between $\approx 9$ and $50$~K, the gas density is between $\approx 75$ and $2.2\times10^{8}$~cm$^{-3}$,
and the visual extinction ranges between $0$ and $62$ magnitudes.
We should note that freeze-out and desorption processes are considered in the chemical modeling,
but are not listed in Table~\ref{tabkey_reac_S}.

The sulfur chemistry in outer disk regions begins with S$^{+}$ reacting with light hydrocarbons such as
CH$_{3}$, CH$_{4}$, C$_{2}$H, C$_{3}$H, producing HCS$^{+}$, HC$_{3}$S$^{+}$, C$_{2}$S$^{+}$ or C$_{3}$S$^{+}$ ions, respectively.
The light hydrocarbons are rapidly formed via gas-phase chemistry of C$^{+}$ and later almost entirely converted to CO or freeze out.
The sulfur-bearing ions dissociatively recombine with electrons, producing smaller neutral sulfurette
molecules. Other slower initial formation routes involve barrierless neutral-neutral reactions with atomic sulfur, e.g.
S + CH, OH, CH$_{2}$, or C$_{2}$, forming CS or SO. At later times, the evolution of
sulfur-bearing species is governed by a closed cycle where protonation processes (via HCO$^{+}$, H$_{3}$O$^{+}$, H$_{3}^{+}$) are equilibrated by the
dissociative recombination reactions, and, for some species, by gas-grain interactions and surface chemistry.
In the following we describe disk chemistry for the individual sulfur-bearing molecules in detail.

The chemical evolution of CS proceeds essentially in the gas phase. CS is
synthesized by dissociative recombination of C$_{2}$S$^{+}$, C$_{3}$S$^{+}$, HCS$^+$, HC$_{2}$S$^+$, and H$_3$CS$^+$,
and by slower neutral-neutral reactions S + CH, CH$_{2}$, or C$_{2}$ and, later, by the SO + C reaction.
The major destruction pathways are freeze-out ($T \lesssim 30$~K), slow neutral-neutral reaction CS + CH $\rightarrow$ C$_{2}$S + H,
charge transfer with ionized hydrogen atoms, and protonation reactions, see Table~\ref{tabkey_reac_S}.

In contrast, the chemical evolution of H$_{2}$S is governed by the surface processes.
The key reaction for H$_{2}$S is the surface hydrogenation of HS followed by desorption. In turn, HS ice can be produced by the
hydrogenation of surface S atoms, or by the freeze-out of the gaseous HS formed via the S + OH reaction.
The major destruction pathways for H$_2$S is freeze-out onto dust surfaces (at $T \lesssim 40-45$~K),
charge transfer reaction with atomic hydrogen, destruction via ion-molecule reaction with C$^{+}$, and protonation reactions.

The chemical evolution of SO and SO$_{2}$ is tightly linked and occurs in the gas and ice phases.
The sulfur monoxide SO is produced by the slow oxidation of S, neutral-neutral reactions between S and OH and O with HS,
and surface reactions between oxygen and S or HS ices. SO is mainly destroyed by exothermic neutral-neutral reactions
with O, C, OH, freeze-out ($T \lesssim 40$~K), charge transfer reactions with atomic hydrogen, and protonation reactions.

Sulfur dioxide SO$_2$ is synthesized by the neutral-neutral reactions between SO and OH, slow radiative
association of SO and O, and the same SO + O reaction occurring on the dust surfaces. The SO$_{2}$
major removal channels are freeze-out at $T\la 60-80$~K, ion-molecule reaction with C$^{+}$ and neutral-neutral
reaction with C, and protonation reactions.

Finally, the chemical evolution of OCS begins with oxidation of HCS and radiative association
of S and CO. OCS can also be produced by the surface recombination of
S and CO as well as CS and O ices. The removal of OCS includes freeze-out ($T \lesssim 45$~K), destruction via ion-molecule reaction with C$^{+}$,
charge transfer reactions with H atoms, and protonation reactions.

\section{Discussion}
\label{sec:diss}

\subsection{Comparison with previous studies of sulfur chemistry}
\label{sec:diss:old_studies}

While many sulfur-bearing species have been routinely observed in low-mass star-forming regions, the previous searches
have found only very few S-bearing species in disks, mostly CS \citep[e.g.,][]{2014prpl.conf..317D}.
One of the first detailed studies of the sulfur chemistry around LkCa~15,
MWC~480, DM~Tauri, and GO~Tauri, using the IRAM~30-m antenna and the gas-grain NAUTILUS chemical code,
have been performed by \citet{2011A&A...535A.104D}.
No H$_{2}$S $1_{1.0}-1_{0,1}$ emission at 168.8~GHz and SO $2_{2,3}-1_{1,2}$ emission at 99.3~GHz were detected.
The comprehensive analysis of the upper limits for SO and the detected CS emission pointed to an elemental C/O ratio of 1.2,
albeit the H$_{2}$S upper limits were
not reproduced, suggesting that either H$_{2}$S remains locked onto grain surfaces or reacts with other species.

Our results agree with these finding, as our best-fit model also requires an elemental C/O ratio of $\gtrsim 1$
in the molecular layer of the DM~Tau disk, see Fig.~\ref{fig:sulfur_chem_C2O}. In this scenario almost all of the oxygen is bound in CO and too little
O is available for synthesis of other O-bearing species, including SO and SO$_{2}$.
We cannot constrain the C/O ratio in the molecular layer more accurately due to the non-detection of SO and
SO$_{2}$ and other S-bearing species and the poorly known abundance of elemental sulfur available for disk chemistry.
Unlike in \citet{2011A&A...535A.104D}, our sensitive upper limit of the H$_{2}$S column density is well explained by
all models except the model with high elemental sulfur abundance or high UV intensity (see Figs.~\ref{fig:sulfur_chem_initial_abundance}
and \ref{fig:sulfur_chem_LISUV}).
Our modeling shows that the H$_{2}$S abundance is not sensitive to other disk physical and chemical properties, see Table~\ref{tabouter_disk_tracers}.
{\it Thus, the combination of CS, SO or SO$_{2}$, and H$_{2}$S lines (or very stringent upper limits) may allow in the future to
more reliably constrain both the C/O ratio and the sulfur elemental abundance in the molecular layers of disks with well-known physical structure.}

In addition, our models can explain the upper limits of the H$_2$S column density,
where the previous study by \citet{2011A&A...535A.104D} failed to reproduce it.
To find the reason for this discrepancy, we first compared the gas-phase and grain surface
networks that have been used in \citet{2011A&A...535A.104D}  studies and
found that they are fairly similar. Next, the gas-phase abundance and
column density of H$_2$S in the outer part of the DM~Tau disk is controlled by the freeze-out and
desorption and thus is sensitive to the adopted disk temperature structure. \citet{2011A&A...535A.104D}
have used a two-layered parametric disk model and considered a representative radius of 300~au.
At that radius, the midplane has a temperature of 10~K an vertical extend $z/H_r \approx 1.5$.
Above the midplane, a molecular layer with a temperature of $\sim 17$~K begins.
In comparison, in our disk model at $r=300$~au the midplane has a temperature of 12~K
and extends vertically up until $z/H_r \approx 1$, after which a warmer molecular layer
begins (with temperatures up to $\sim 30-40$~K).
Finally, we present the median H$_2$S column densities over the outer $100-800$~au region.
In fact, in our DM~Tau model the H$_2$S column densities increase toward the outer edge of
the disk, and at $r=300$~au the typical values of $N($H$_2$S$)$ are about $10^{13}$ - a few
times $10^{13}$~cm$^{-2}$. These values are comparable with the values at $r=300$~au
calculated by \citet{2011A&A...535A.104D}, given the modeling uncertainties of a factor of several.

\citet{Guilloteau_ea13} have performed a chemical survey of 42 T~Tauri and Herbig~Ae systems
located in the Taurus-Auriga region with the IRAM 30-m antenna (1~hour/source integration), including the SO emission line.
They have detected SO in 7 sources, which also showed strong H$_{2}$CO emission. The observed SO line profiles suggest that the emission is coming from outflows or envelopes rather than from Keplerian rotating disks.
Later, \citet{2016A&A...592A.124G} have performed a deeper survey of 30 disks in Taurus-Auriga with the IRAM 30-m antenna
($>1-2$~hours/source integration), targeting two SO lines. In addition to many CS detections,
they have detected SO in four T~Tauri and Herbig~Ae sources, which traces predominantly outflow/shocked regions.

The SO molecule has also been detected and imaged in the transitional disk around the Herbig~A0 star AB~Aur by \citet{PachecoVazquez:2016fc}.
They found that the SO emission has a ring-like shape and that SO is likely depleted from the gas phase in the
horseshoe-shaped dust trap in the inner disk.
Recently, \citet{Booth_ea17_SO} have detected two SO 7-6 emission lines in the HD~100546 disk at around 300~GHz and found that the SO emission does not
come solely from the Keplerian disk, but likely traces either a disk wind, an inner disk warp (seen in CO emission) or an accretion shock from a circumplanetary disk associated with the proposed protoplanet embedded in the disk at 50~au.

Indeed, SO and SO$_{2}$ are widely used as shock tracers in the studies of high-mass star-forming regions or low-mass Class~0 and I protostars
and bipolar outflows associated with the low mass stars
\citep{1993MNRAS.262..915P,2010A&A...522A..91T,2014A&A...567A..95E,2014A&A...563A..97G}. Recently, \citet{2015A&A...581A..85P}
have observed SO emission rings around the Class~0 protostars L1527 and HH212, which likely trace shocked gas at the disk-envelope interface.
As we have shown in Section~\ref{sec:analysis}, SO and SO$_{2}$ are partly produced via dust surface process and remain frozen at $T \lesssim 45-80$~K,
while CS is produced in the gas phase and is more volatile, with evaporation temperature of $\sim 30$~K. Thus, shocks or elevated temperatures
available in the inner, $r \lesssim 30-100$~au regions of T~Tauri disks are required to effectively put SO and SO$_{2}$ ices back into the
gas phase, while in the outer disks SO and SO$_{2}$ desorption is less efficient and driven by CRP-induced or interstellar UV photons.
Apparently, this is not the case for the quiescent DM~Tau disk (or, at least, its outermost regions).

\subsection{The non-solar C/O ratios in disks as traced by CS/SO}
\label{sec:diss:C_O_ratios}

Our main finding is that the observed data are best explained by a C/O ratio greater than 1.
This agrees well with the other recent observational results.
\citet{Bergin:2016ge} have observed C$_{2}$H in both TW~Hya and DM~Tau with ALMA, and also found cyclic C$_{3}$H$_{2}$ in TW~Hya.
These hydrocarbons show bright emission rings arising at the edge of the mm-dust disk, which
can only be reproduced in the chemical models with a C/O ratio $>1$ and a strong UV field in molecular layers, as in our study.
They explained the effect of non-solar C/O ratios in the molecular layer by predominant removal of oxygen in form of water ice into the
dark and cold midplane by the sedimentation of large, pebble-sized dust grains. This leads
to a stratified C/O structure in the disk with higher ratios within the UV-dominated upper layer and solar-like C/O ratios in the
ice-rich disk midplane.

A similar result was obtained by \citet{Kama_ea16}, who have observed of the CI, OI, C$_{2}$H, CO and CII lines
to constrain the carbon and oxygen abundances in TW~Hya and HD~100546. To match the observations, a strong depletion of the
C and O reservoirs in the cold T~Tauri disk around TW~Hya is required, but the warmer disk around HD~100546 is much less depleted.
Both disks however needed a model with C/O $>1$ in order to match the observations. The strong C and O depletion in TW~Hya
has also been confirmed by the CO isotopologue observations and detailed modeling of \citet{Schwarz:2016ch}.
\citet{McClure_ea16} have also found a moderate C and O depletion in DM~Tau and a strong depletion of the
C and O reservoirs in GM~Aur.
The authors proposed a mechanism of locking the refractory O-rich ices (like H$_{2}$O) by the sedimentation of grains into the
dark midplane where it cannot easily desorb back into the gas phase.

This mechanism of volatile depletion in disks has been investigated by \citet{Du:2015kz} and \citet{Krijt_ea16}.
\citet{Du:2015kz} have proposed that the elemental C and O abundances can be reduced in the upper layers of the outer disk as major
C- and O-bearing volatiles reside mainly at the midplane, locked up in the icy mantles of mm-sized dust grains partly decoupled from
the gas, which hence cannot be transported upward by turbulence. \citet{Krijt_ea16} have developed a dynamical model of evolving dust grains
in a disk. They found that dust coagulation enhanced by the presence of ice mantles benefits dust growth and hence dust vertical settling,
leading to a strong depletion of water in the disk atmosphere, where the gas-phase C/O becomes $\sim 1$ and higher.
This removal mechanism is supported by the previous dynamical studies of evolving
dust grains in disks and laboratory experiments \citep[e.g.,][]{Birnstiel_ea10a,2015ApJ...815..109P,2016ApJ...827...63M}.

An alternative explanation has been proposed by \citet{2013ApJ...776L..38F}. They have obtained a low disk-averaged gas-phase CO abundance in TW~Hya
and inferred that this could be due to chemical destruction of CO by the X-ray-produced He$^{+}$ ions, followed by rapid formation of heavy carbon chains
that freeze-out onto dust grains, while oxygen goes into the synthesis of other species, like CO$_{2}$, water, or complex organic molecules.
In our disk chemical model the destruction of CO by the X-ray-produced He$^{+}$ ions occurs only in the upper layers at $r \lesssim 3-10$~au
and hence cannot explain the derived C/O ratio of $\gtrsim 1$ from our sulfur data probing the outer, $>100$~au region of the DM~Tau disk.

\subsection{Sulfur chemistry as tracer of water snowline in disks}
\label{sec:diss:snowline}

A defining feature for planet formation is the water
snowline that shapes the properties of emerging planets
\citep{2016SSRv..205...41B}. It is temperature-, pressure-, and composition- dependent,
and has a value of $T \sim 140-160$~K, which
corresponds to the radial distances of $\sim 1-3$~au in a typical T~Tauri disk
\citep{2014prpl.conf..835V}. Condensation of water is connected to the C/O ratio of gas
and ices in protoplanetary disks \citep{Bergin:2015in,Oberg:2016fh}. Despite its importance to protoplanetary disk evolution and planet formation, a direct detection of the water snowline in protoplanetary disks is still missing.

\begin{figure}
 \begin{center}
\includegraphics[]{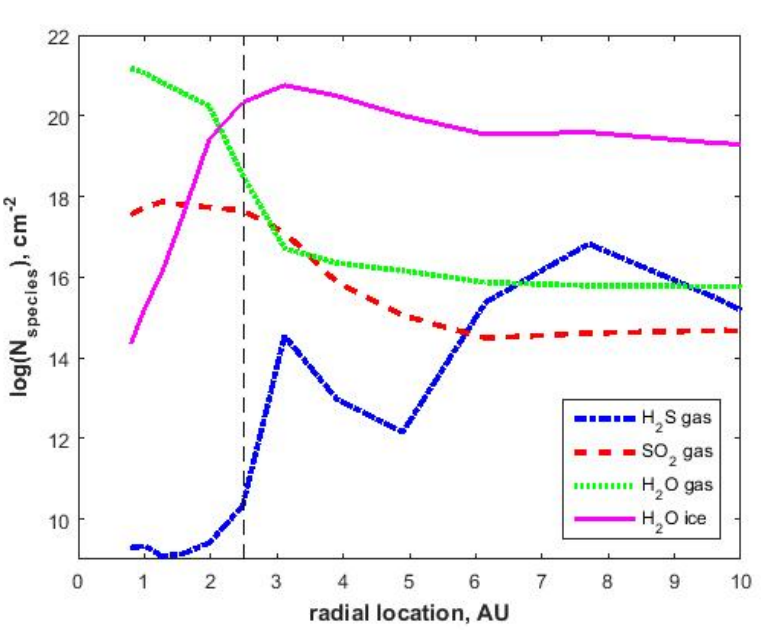}
\caption{Our DM~Tau chemical model suggests a relationship between the column densities of H$_{2}$S, SO$_{2}$,
and H$_{2}$O. The black vertical line marks the snowline location at $\sim 2.5$.}
\label{fig:snowline}
\end{center}
\end{figure}

While the chemical modeling of the DM~Tau disk presented above was mainly focussed on the interpretation of the outer-disk averaged sulfur chemistry, our model shows another interesting result, namely, a close relationship between the column densities of H$_{2}$S, SO$_{2}$, and H$_{2}$O in the inner disk region at $<10$~au (Fig.~\ref{fig:snowline}).
Specifically, a three order of magnitude rise in the column density of SO$_{2}$ is observed starting at $\sim 4-5$~au, which is at a
distance about twice larger than the snowline location of
$\sim 2-2.5$~au. The rise in SO$_{2}$ prior and inside the snowline is caused by the faster surface synthesis due to increased
grain temperatures and more rapid thermal hopping and desorption of water ice that brings oxygen back to the gas phase and
enables faster SO$_{2}$ gas-phase chemistry.
H$_{2}$S shows the opposite trend, with a rapid decrease in column density of six orders of magnitude inside
the snowline. This is due to the fact that H$_{2}$S forms efficiently on dust grains,
which is effective only at low temperatures of $\lesssim 20-40$~K and limited by hydrogen residence time on grains. At higher temperatures sulfur surface chemistry begins to produce heavier O-bearing sulfur species like SO$_{2}$ and OCS.
Potentially, these chemical trends can be observed in inner disk regions with the forthcoming JWST mission.

\section{Conclusions}
\label{sec:concl}
We have performed a new search for the sulfur-bearing molecules CS, SO, SO$_{2}$, OCS, CCS, H$_{2}$CS and H$_{2}$S in the DM~Tau disk with ALMA
in Cycle~3. We have detected CS and have a tentative SO$_{2}$ detection. Failure to detect sulfur-bearing molecules in DM~Tau other than CS
implies either efficient freeze-out of H$_{2}$S, SO, SO$_{2}$, OCS or that this sulfur is bound in other species,
possibly refractory compounds as identified in comets.
For the data analysis, we have merged
the new CS (3-2) ALMA Cycle 3 data with our previous PdBI data and stacked the two transitions for SO, SO$_{2}$ and OCS.
We have employed a 1D non-LTE radiative transfer code RADEX to obtain the disk-averaged CS column densities and the upper limits for the other sulfurette
molecules. Also, we used a 1+1D forward-modeling tool DiskFit to derive the corresponding radial distributions.
Both these methods show that the CS column density is $\sim 2-6 \times 10^{12}$~cm$^{-2}$ (with a factor of two uncertainty).

These observed values have been compared with a suite of detailed physical-chemical models of the DM~Tau disk by varying key parameters such as
the elemental sulfur abundance, grain sizes, amount of high-energy radiation, turbulent mixing, C/O ratios, surface chemistry efficiency.
We have found that the observed data and, in particular, the CS/SO ratio of $\gtrsim 1$, can only be reliably explained by a disk model with a C/O
ratio of $\gtrsim 1$, in agreement with the previous studies of sulfur content of protoplanetary disks. This is also consistent with the recent observations of bright hydrocarbon emission rings and non-solar C/O $\gtrsim 1$ ratios found in DM~Tau, TW~Hya and a few other disks.
The stronger depletion of oxygen-bearing S-species compared to CS is likely linked to the proposed removal mechanism of oxygen from
the disk upper layers by growing and sedimenting ice-coated grains.
Furthermore, our chemical modeling demonstrated that sulfur-bearing species, if more would be detected in disks, could be robust tracers of other
key disk properties, such as efficiency of surface reactions, grain sizes, and amount of high-energy radiation that is able to reach the
disk molecular layers.
Thus more sensitive observations of sulfur-bearing molecules in the disks with well-studied structures are required with
ALMA and NOEMA facilities.

\begin{acknowledgements}
This paper makes use of the following ALMA data: ADS/JAO.ALMA\#2015.1.00296.S. ALMA is a partnership of ESO
(representing its member states), NSF (USA) and NINS (Japan), together with NRC (Canada) and NSC and ASIAA
(Taiwan) and KASI (Republic of Korea), in cooperation with the Republic of Chile. The Joint ALMA Observatory
is operated by ESO, auI/NRAO and NAOJ. This work was supported by the National Programs PCMI and PNPS from INSU-CNRS.
DS acknowledges support from the Heidelberg Institute of Theoretical Studies for the project
''Chemical kinetics models and visualization tools: Bridging biology and astronomy''.
DF and CF acknowledge support from the Italian Ministry of Education, Universities and Research, project SIR (RBSI14ZRHR).
This research made use of NASA's Astrophysics Data System. The figures in this paper were
constructed with the \textsc{matplotlib} package \citep{Hunter2007} and \textsc{aplpy} package hosted at \url{http://aplpy.github.io}.
\end{acknowledgements}

\bibliographystyle{aa}

\end{document}